\documentclass[aps,prab,reprint,superscriptaddress,showpacs,showkeys,letter]{revtex4-1}
\usepackage{graphicx}

\usepackage{dcolumn}
\usepackage{bm}
\usepackage{amsmath}
\usepackage[caption=false]{subfig}
\usepackage[none]{hyphenat}
\usepackage{color}
\usepackage{mathrsfs}
\usepackage{epstopdf}
\usepackage{amssymb}
\usepackage{mathrsfs}

\begin{document}


\title{Structure resonance crossing in space charge dominated beams}

\author{Zhicong Liu}
\affiliation{%
Key Laboratory of Particle Acceleration Physics and Technology, Institute of High Energy Physics, Chinese Academy of Sciences, 19(B) Yuquan Road, Beijing 100049, China}
\affiliation{%
University of Chinese Academy of Sciences, Beijing 100049, China}

\author{Chao Li}
\email{lichao@ihep.ac.cn}
\affiliation{%
Key Laboratory of Particle Acceleration Physics and Technology, Institute of High Energy Physics, Chinese Academy of Sciences, 19(B) Yuquan Road, Beijing 100049, China}
\affiliation{%
Institut f\"{u}r Kernphysik--4, Forschungszentrum J\"{u}lich, 52425 J\"{u}lich, Germany}

\author{Qing Qin}
\affiliation{%
Key Laboratory of Particle Acceleration Physics and Technology, Institute of High Energy Physics, Chinese Academy of Sciences, 19(B) Yuquan Road, Beijing 100049, China}


\begin{abstract}
As an extension of previous theoretical study on the coherent structure resonance
due to space charge effects [Chao Li and R. A. Jameson, Phys. Rev. Accel. Beams 21, 024204, 2018],
this paper aims to demonstrate how the beam, as a whole, is spontaneously affected
when the predicted mixed coherent 2nd/4th order structure resonance stop band
around $90^{\circ}$ phase advance is crossed.
The beam characteristics during the structure resonance crossing,
such as the rms emittance growth and the appearance of 2-fold/4-fold structure
in phase space, are well explained by the mixture characteristic of different orders of structure resonance. The related ``attracting''
and ``repulsive'' effects in the structure resonance stop band
from below and above crossing are considered as a natural beam reaction
to the coherent structure resonance that the beam spontaneously moves to a structure resonance 
free region then the space charge takes a weaker importance. In the PIC simulation, it is found
that the emittance growth is positively related to the time that
the beam spends inside the structure resonance stop band. As a potential candidate
mechanism, the incoherent particle-core resonance also has been checked.
It is found that this incoherent resonance has basic
difficulties in explaining the results obtained from the self-consistent PIC simulation. The
incoherent particle-core resonance might lead to phase space distortion,
emittance growth, or beam halo formation only on a long-time scale.
This clarity of the discrepancies between coherent and incoherent resonance mechanisms will lead to a better
understanding of the  numerical study and  experimental researches obtained recently. In addition, similar understanding
can be extended to the study of higher order structure resonance.
\end{abstract}


\maketitle

\clearpage

\section{Introduction}

The nonlinear beam dynamics in accelerators has been studied with analytical,
numerical and experimental approaches for several decades~\cite{1,2,45}.
Nowadays, the resonances driven by various nonlinear effects in
accelerators are considered as the main sources leading to beam
deterioration such as rms emittance growth, beam halo, and beam losses.
Generally, the studies on beam dynamic behavior are in the frame of the
Hamiltonian system and usually the nonlinear terms in the complex particle Hamiltonian
 are treated as perturbations. The resonance conditions are
obtained with linearized perturbation techniques. Besides normal static external elements,
the nonlinearity from the internal space charge effect
is another main source that must be considered and treated carefully. Moreover,
the nonlinear space charge couples the motions in different degrees of freedom in a self-consistent
manner, especially in low energy and high current machines~\cite{3}.

According to the beam motions to be focused on, the descriptions of nonlinear
resonance are divided into single particle dynamics
level (incoherent effect) and rms beam dynamics level (coherent effect).
A typical example of single particle dynamics description is the tune diagram,
which is filled with various resonance lines, $n\nu_x+m\nu_y=l$,
due to the multipoles and lattice imperfection~\cite{1}.
It is widely used as a guidance on working point selection in the 
designs and operations. Considering the nonlinear effect,
a wide tune spread will be formed~\cite{cousineau2006space}.
When a beam is centered near the resonance lines in tune diagram, some
particles can periodically cross the resonances
if the synchronous motion is taken into account ~\cite{5}.
As to the coherent effect (rms level), nonlinearity from external elements
and internal space charge effect plays a key role
in beam collective instabilities \cite{sacherer1968transverse, 7},
which normally requires solving the perturbed Vlasov equation
in a self-consistent manner \cite{8,9,10}.
However, the analytical solution of such system is not trivial, and the
solvable problems are still limited to certain specific cases. Great efforts
have been paid to extend these models to cover problems with various beam
conditions.

As to the study of collective instability due to space charge effect
with ion beams, one branch starts from the rms envelope equations~\cite{13},
with which the 2nd order (envelope instability) structure resonance is well studied ~\cite{14,15,16,17,21,22},
while another branch is to solve the Vlasov-Poisson equations self-consistently~\cite{li16collective,12,18,19}.
Recently, a general theoretical study of the Vlasov-Poisson model in space charge physics
is given in Ref.~\cite{li16collective,12}, where the resonance phenomenon
discussed in Ref.~\cite{18} has been extended to cover problems with
various initial beam distributions and focusing conditions.
It is noteworthy that the driving force of the resonance derived from the
Vlasov-Poisson does not come from the rms mismatch but from inner density mismatch.
In Ref. \cite{12}, it states that the ``structure resonance"
can take place among these constructed eigenmodes if the lattice parameters are
not elaborately optimized; it also proves the lower order structure resonance stop
band can be naturally treated as one component of the higher order structure
resonance stop bands; in addition, the 2nd order even structure resonance exactly describes the same
coherent structure resonance as those from the envelope dynamics.

In fact, the external focusing strength usually changes as the beam being accelerated,
which possibly results in the ``structure resonance crossing". This paper will
focus on how the coherent and incoherent characteristics of a initial rms matched beam and inner particles
are spontaneously affected during the beam crossing the structure resonance stop band. As an
extension of the theoretical study of coherent structure resonance with Vlasov-Poisson model \cite{li16collective, 12},
the 2nd order stop band around $90^{\circ}$ phase advance is chosen for detailed study to verify
the validity of the theoretical prediction in the simulation and demonstrate
the transient interaction between beam and structure resonances. These studies can
be  extended to higher order  structure resonances with phase advance around $60^{\circ}$
~(3rd order structure resonance) or $120^{\circ}$ (6th order structure resonance)
\cite{36,40}.  It is noteworthy that the fact that lower order coherent structure resonances are components of higher order
structure resonances is the key to understand the results obtained from simulation and
experiment reported in Ref~\cite{32,33}.

The incoherent particle-core resonance, which is considered as one of the potential
mechanism for the generation of particle tail in phase space, is intuitively used to
explain the tail structure in phase space \cite{25}. By simulating with initially rms matched beam conditions,
we show that it has some difficulties to explain the basic phase space structures. The incoherent resonance
may affect the beam but only on a long time scale.

In Section \ref{section:model}, the model of structure resonance is briefly introduced. The
terminologies used to describe the coherent and incoherent effect are introduced.
In Section \ref{section:coherent}, the equations to get the stop band are given explicitly. The
phenomena when the beam crosses the coherent structure resonances are simulated
with multi-particle PIC code and appropriately discussed in a transient sense.
In Section \ref{section:incoherent}, the incoherent particle characteristic is discussed. The
conclusion and summary are given in Section \ref{section:summary}.

\section{Physical model and terminologies}  \label{section:model}
In this section, we briefly introduce the physical model and basic approach to deal with
structure resonance. The solvable coupled Vlasov-Poisson equation is limited to
the 4D KV distribution assumption~\cite{20}.
In the following, the periodic FODO channel is used to model the evolution of
coasting beam in accelerators. Considering the fact that the Hamiltonian of a
matched beam in a periodic focusing channel is conserved, the equilibrium
distribution function can be expressed as a function of generalized
Hamiltonian:
\begin{eqnarray}\label{eq2.1}
  &&f_0(x,p_x,y,p_y) = f(H_0),  \\
  &&H_0 = k_x(s)x^2+p_x^2+k_y(s)y^2+p_y^2+V_{sc}(x,y),
\end{eqnarray}
where $k_x(s)$ and $k_y(s)$ are the external focusing strength supplied by the
quadrupoles, and $V_{sc}(x,y)$ is the space charge potential. The distribution function $f_0$ must meet the Vlasov equation and Poisson's equation
\begin{eqnarray}  \label{eq2.2}
  &&\frac{\partial f_0}{\partial s} + [f_0,H_0] = 0,   \\
  &&\Delta V_{sc}(x,y) = \frac{1}{\epsilon_0}  \int \int f_0 dxdy,
\end{eqnarray}
where $[,]$ is the Poisson bracket operator. Assuming that there exists a
perturbation $f_1$ on the particle distribution function, it will lead to a perturbed space
charge potential $V_1=H_1$. Thus, the first order linearized Vlasov and Poisson
equation can be obtained as:
\begin{eqnarray}\label{eq2.3}
  &&\frac{\partial f_1}{\partial s} + [f_1,H_0] +  [f_0,V_1]=0,   \\
  &&\Delta V_1(x,y) = \frac{1}{\epsilon_0}  \int \int f_1 dxdy.
\end{eqnarray}

The solvable sets of the equations are limited to the assumption of ideal KV
distribution $f_0=\delta(H_0)$. In general, the perturbed space charge
potential can be expressed as the form of polynomial inside the beam
\begin{eqnarray}\label{eq2.4}
V_1=\sum_{m=0}^{n}A_m(s)x^{n-m}y^m+\sum_{m=0}^{n-2}A_m^{(1)}(s)x^{n-m-2}y^m+\cdots \nonumber  \\.
\end{eqnarray}
For a given order $n$ in Eq.~(\ref{eq2.4}), the even structure resonance and the
odd structure resonance, which directly represent the tilts of the beam elliptical distribution
 in real space, could be treated separately on the basis of whether the index
$m$ is restricted to even or odd integer values.

Inheriting the terminologies defined in the former researches \cite{li16collective,12},
we briefly introduce the physical meaning of the notations used in the following study.
The collective modes $I_{j;k,l}(s)$ physically represents the integral
of the surface electric field discontinuity from period to period.
Noting $S$ as the length of one focusing period, $I_{j;k,l}(s)$ will meet
$I'(s) = M(S)I(s)$, which is exactly the $Mathieu$ $Equation$.
The stability of the system is decided by the eigenvalues $\lambda$ of the Jacobi matrix $M(S)$.
The phase advance of  $I_{j;k,l}(s)$ is noted as $\Phi_{j;k,l}$;
$\Phi_{e}$ is used to depict the phase advance of the rms matched envelope oscillation characteristics in one period,
which is always $360^{\circ}$; $\Phi_{e}$ naturally represents the periodicity of the
lattice, and the words ``envelope oscillation period" and ``lattice
period" are considered to be equivalent.  $\sigma_s$ is used to describe the single particle phase advance. The nonlinear
effects from external elements and internal space charge cause different particles to have different
particle phase advance $\sigma_s$, leading to a beam phase advance spread.  $\sigma_0$ and $\sigma$ are standard notations
used to evaluate the average focusing strength in one focusing period without and with space charge.
The coherent structure resonance conditions are expressed explicitly as
$\Phi_{j;k,l}^{(1)}+\Phi_{j;k,l}^{(2)}=n\times 360^{\circ}$ and $\Phi_{j;k,l}/\Phi_{e}=n/m$,
It will be shown in the following that the structure
resonance takes place accompanied by the eigenphase locking. The
parameter space where the structure resonance takes place is named as unstable
stop bands. The incoherent particle-core resonance is express as  $\sigma_s/\Phi_{e}=n/m$.
More mathematical details can be found in
Ref.~\cite{li16collective,12,18,19}.

\section{The structure resonance stop band crossing -- coherent effect}  \label{section:coherent}
In the accelerator designs and operations, the principle  ``the resonance
should be passed through as fast as possible if it cannot be avoided'' is widely used in accelerator physics design \cite{li2013ADS,henderson2014SNS}.
In the following, the 2nd order structure resonance around the
$90^{\circ}$ phase advance
is used to demonstrates the interaction between beam and structure resonance
when the 2nd order stop band is crossed on purpose.
As discussed, the 2nd order collective structure resonance is actually one component of
the 4th order collective structure resonance \cite{12}. In the following, the 2nd order structure resonance
particularly refers to this mixed stop band.
For simplicity, the beam
with equal emittance in two degrees of freedom is adopted to model the beam
behavior in symmetric periodic FODO channels~($|k_x|=|k_y|$).

\subsection{The 2nd order collective structure resonance}
For the 2nd order structure resonance, the related perturbed space charge potential inside
near the beam boundary is $V_{2e}=A_0(s)x^2+A_2(s)y^2$ for the even structure resonance and
$V_{2o}=A_1(s)xy$ for the odd structure resonance. Inheriting the notation in
Ref.~\cite{li16collective, 12, 18}, dynamic system $I'(s)=M(s)I(s)$ is constructed with
$(I_{0;2,0}, I_{2;0,2})$  and $(I_{1;1,1}, I_{1;1,-1})$, representing the 2nd
order even and odd structure resonance respectively. The explicit forms of the eigenphases with and
without beam current, structure resonance driving terms,
and the related forms of the perturbed space charge potential can be found in Tab. I in Ref. \cite{12}.
The Jacobi matrix $M(s)$ is with the form
\begin{eqnarray}\label{eq2.5}
M(s)=
\left(
  \begin{array}{cccc}
    0         & 1         & 0         & 0         \\
    J_{21}(s) & J_{22}(s) & J_{23}(s) & 0         \\
    0         & 0         & 0         & 1         \\
    J_{41}(s) & 0         & J_{43}(s) & J_{44}(s) \\
  \end{array}
\right).
\end{eqnarray}
The explicit form of each element $J_{ij}(s)$ in above equation can be found in the Appendix.

\begin{figure*}
    \centering
    \subfloat[\label{sfig:stopbandEVEN}]{
        \includegraphics[width=.48\linewidth]{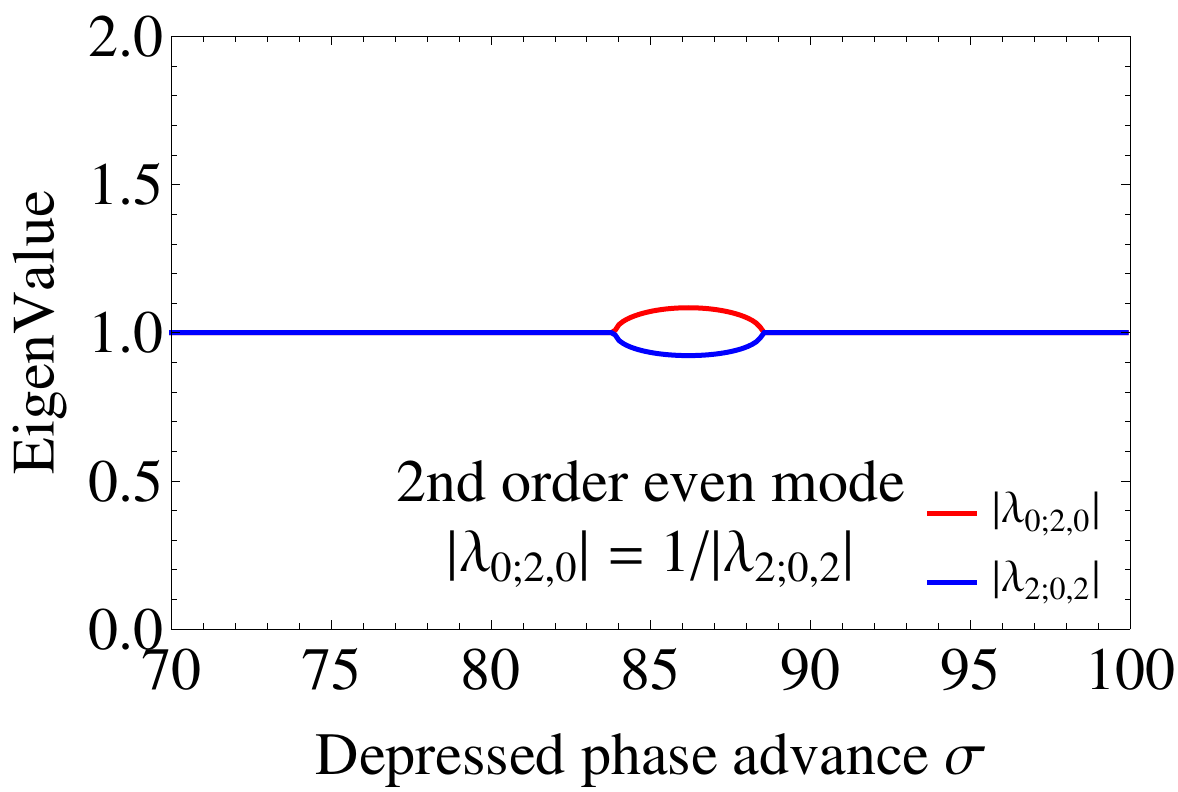}
    }
    \subfloat[\label{sfig:stopbandODD}]{
        \includegraphics[width=.48\linewidth]{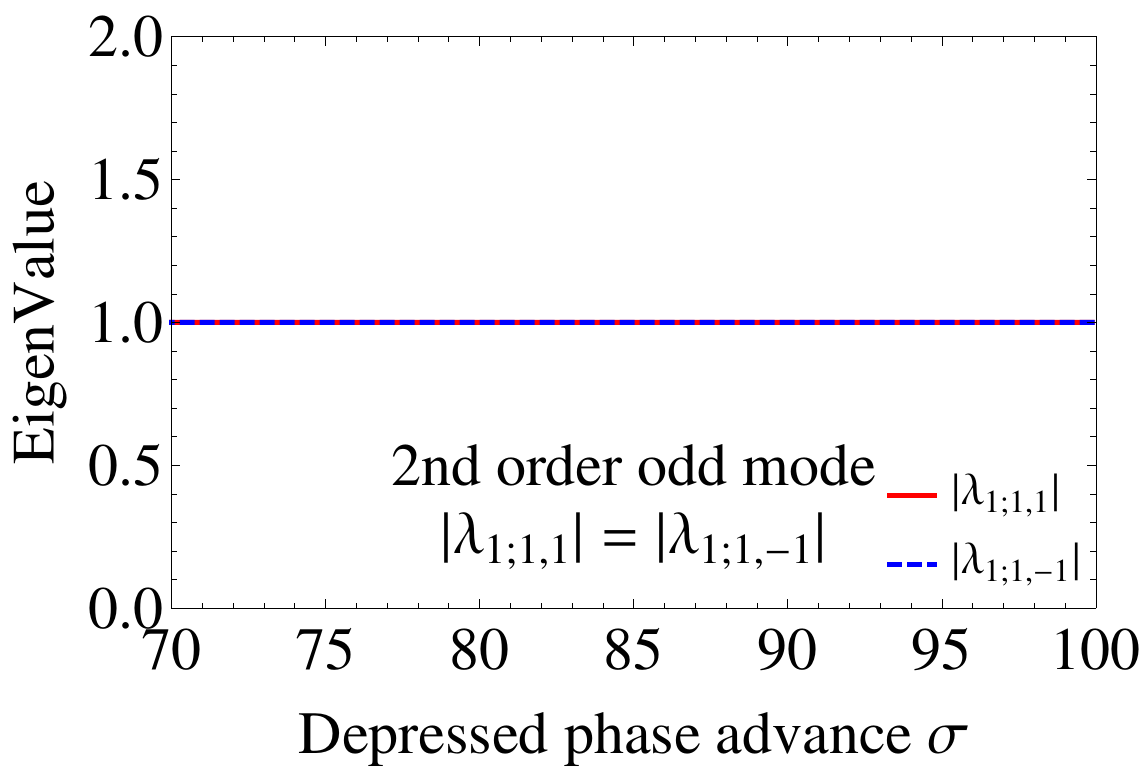}
    }

    \subfloat[\label{sfig:stopbandODD}]{
        \includegraphics[width=.48\linewidth]{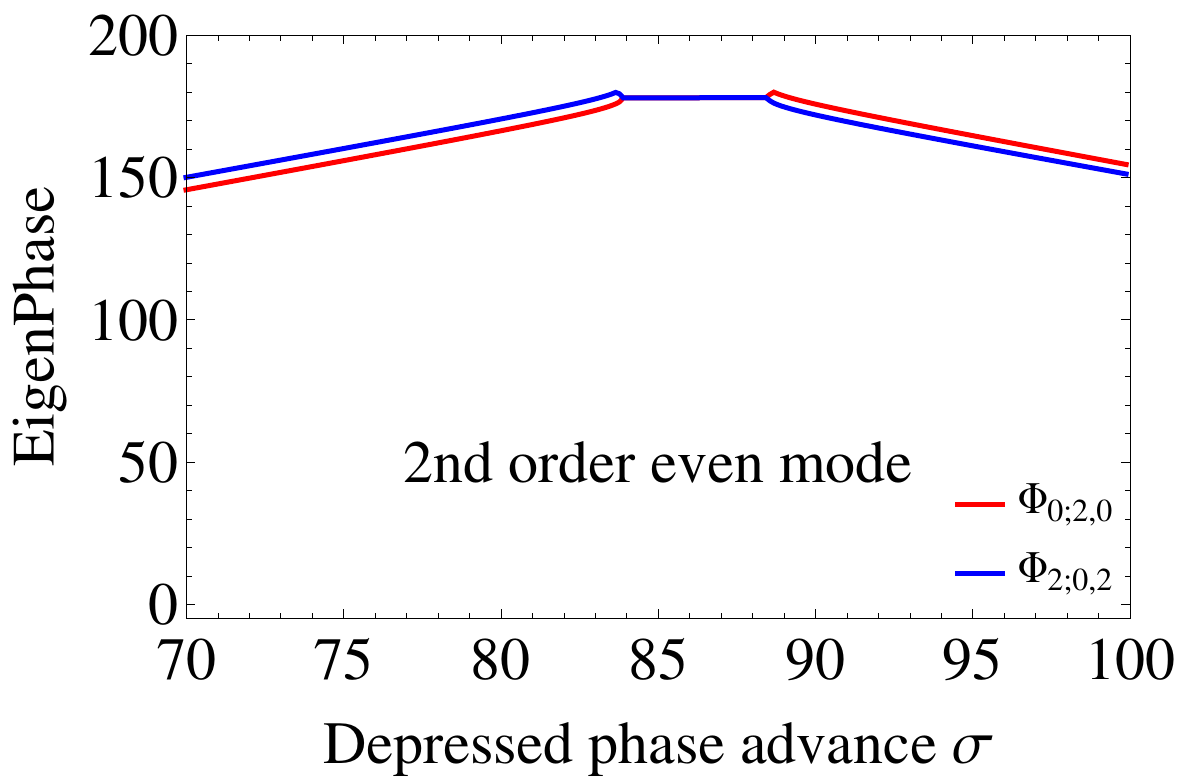}
    }
    \subfloat[\label{sfig:stopbandODD}]{
        \includegraphics[width=.48\linewidth]{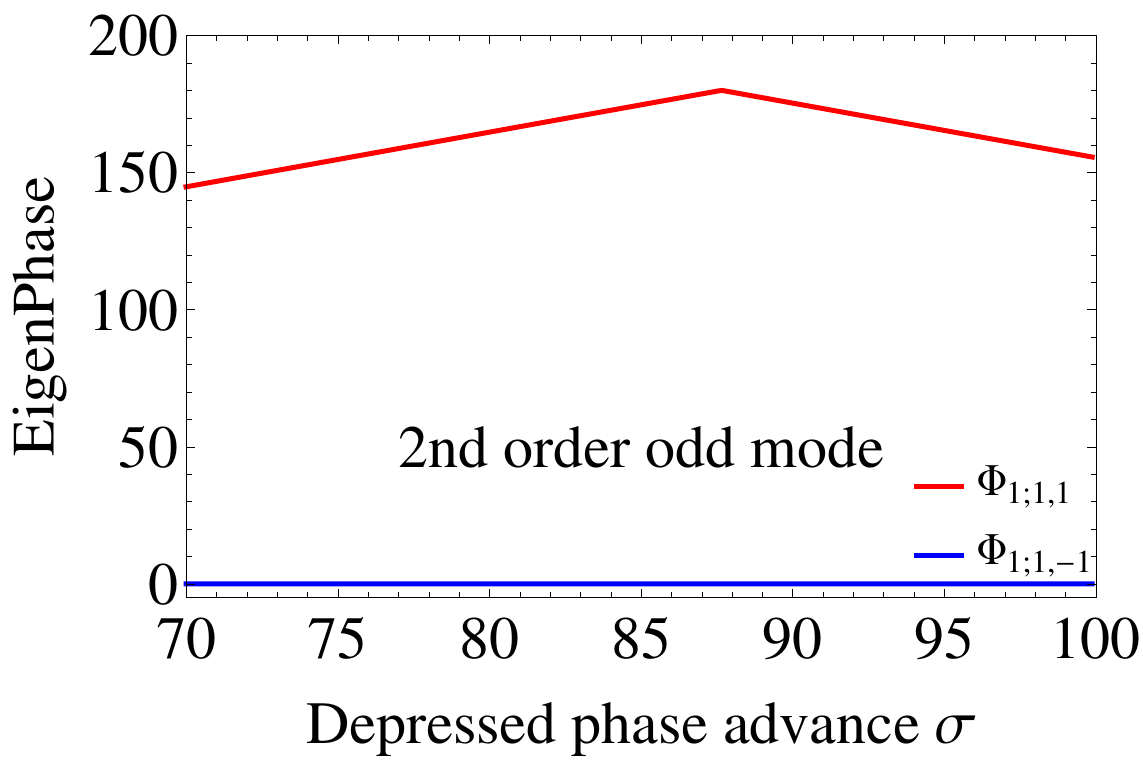}
    }
    \caption{The evolution of absolute eigenvalues $Abs[\lambda_{j;k,l}]$ (above) and eigenphases $\Phi_{j;k,l}$ (below) as a function of the depressed phase advance $\sigma$ for 2nd order even (left) and odd structure resonance (right).}
    \label{fig:stopband}
\end{figure*}

It is noteworthy that the well-known envelope instability, starting from the rms envelope model,
describes the same physics as the 2nd order even structure resonance
\cite{li16collective,12,18}. Recently, the sum and difference instabilities are studied in
Ref.~\cite{21} with Chernin's  model \cite{22}, which gives exactly the same
result as that from the 2nd order odd structure resonance. From the experiment
point of view, great efforts have been put into the spectrum measurement of the
$I_{0;2,0}$ and $I_{2;0,2}$ in GSI~\cite{singh2014observations} and
CERN~\cite{cernAdrian}.

With the help of Eq.~(\ref{eq2.5}),  Fig.~\ref{fig:stopband} shows the eigenvalues
$Abs[\lambda_{j;k,l}]$ and eigenphases $\Phi_{j;k,l}$ of the 2nd order even and
odd structure resonance when the beam current is fixed and the depressed current phase
advance $\sigma$ varies from $100^{\circ}$ to $70^{\circ}$.
Correspondingly, the zero-current phase advance $\sigma_0$ varies roughly from
$110^{\circ}$ to $80^{\circ}$.
Since the symmetric condition chosen here, the phase advances are
equal in two degrees of freedom with and without space charge.
The results are shown in the Fig. 1. For the 2nd order even
structure resonance, whenever the eigenphases merging~(eigenphases locking between
$\Phi_{2;0,2}=\Phi_{0;2,0}$, and it is also termed as confluent resonance \cite{12})
takes place, as region  ($84^\circ$, $89^\circ$), the absolute eigenvalue
$Abs[\lambda]$ leaves the unit 1, representing collective instabilities
and resulting in rms emittance growth. The 2nd order odd structure resonance does not lead to
any instability since the symmetric condition is chosen here.

\subsection{The 2nd order structure resonance crossing study}

\begin{figure*}[thbp]
    \centering
    \subfloat[\label{sfig:75_95phase}]{
        \includegraphics[width=.45\linewidth]{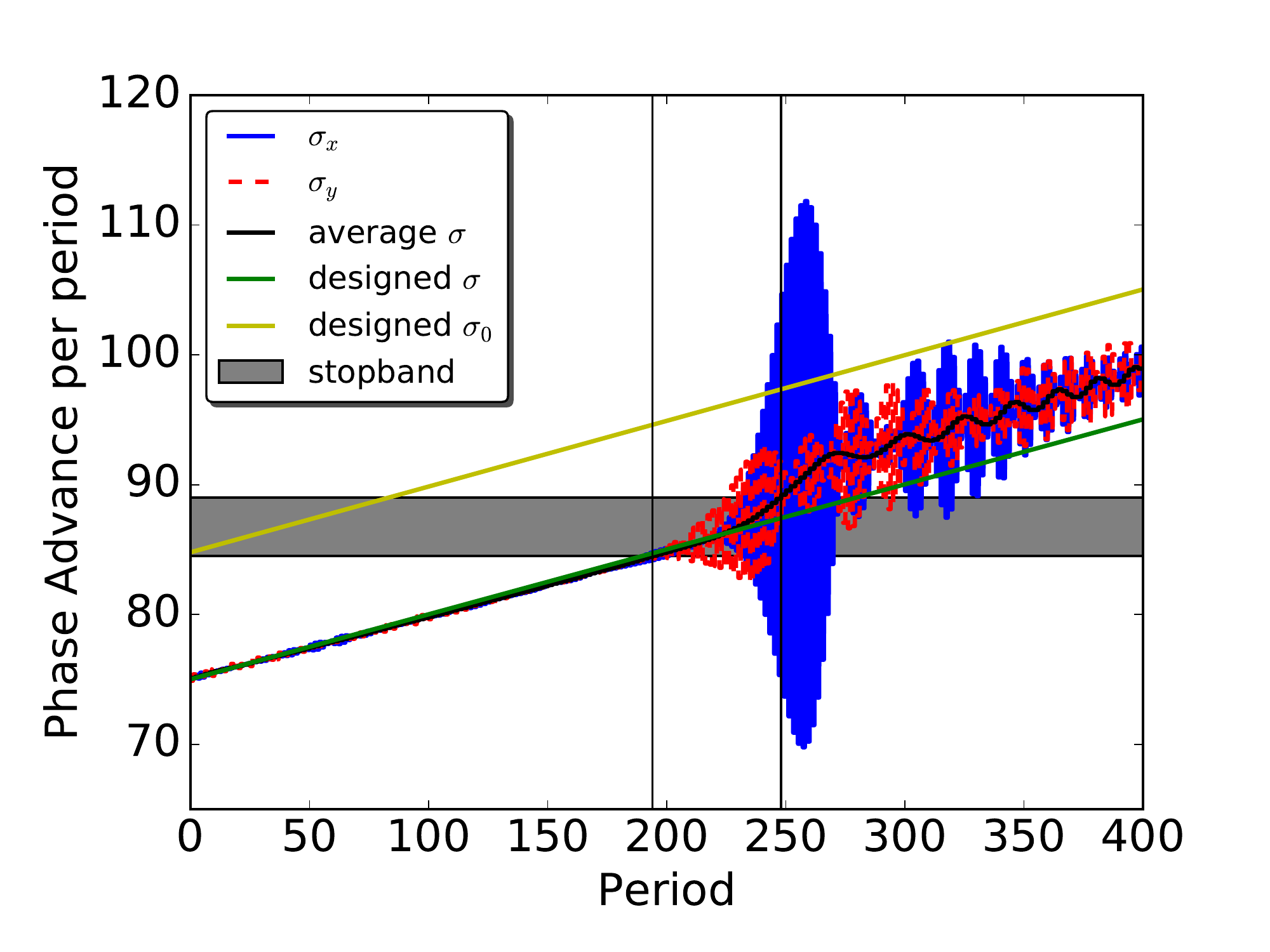}
    }
    \subfloat[\label{sfig:95_75phase}]{
        \includegraphics[width=.45\linewidth]{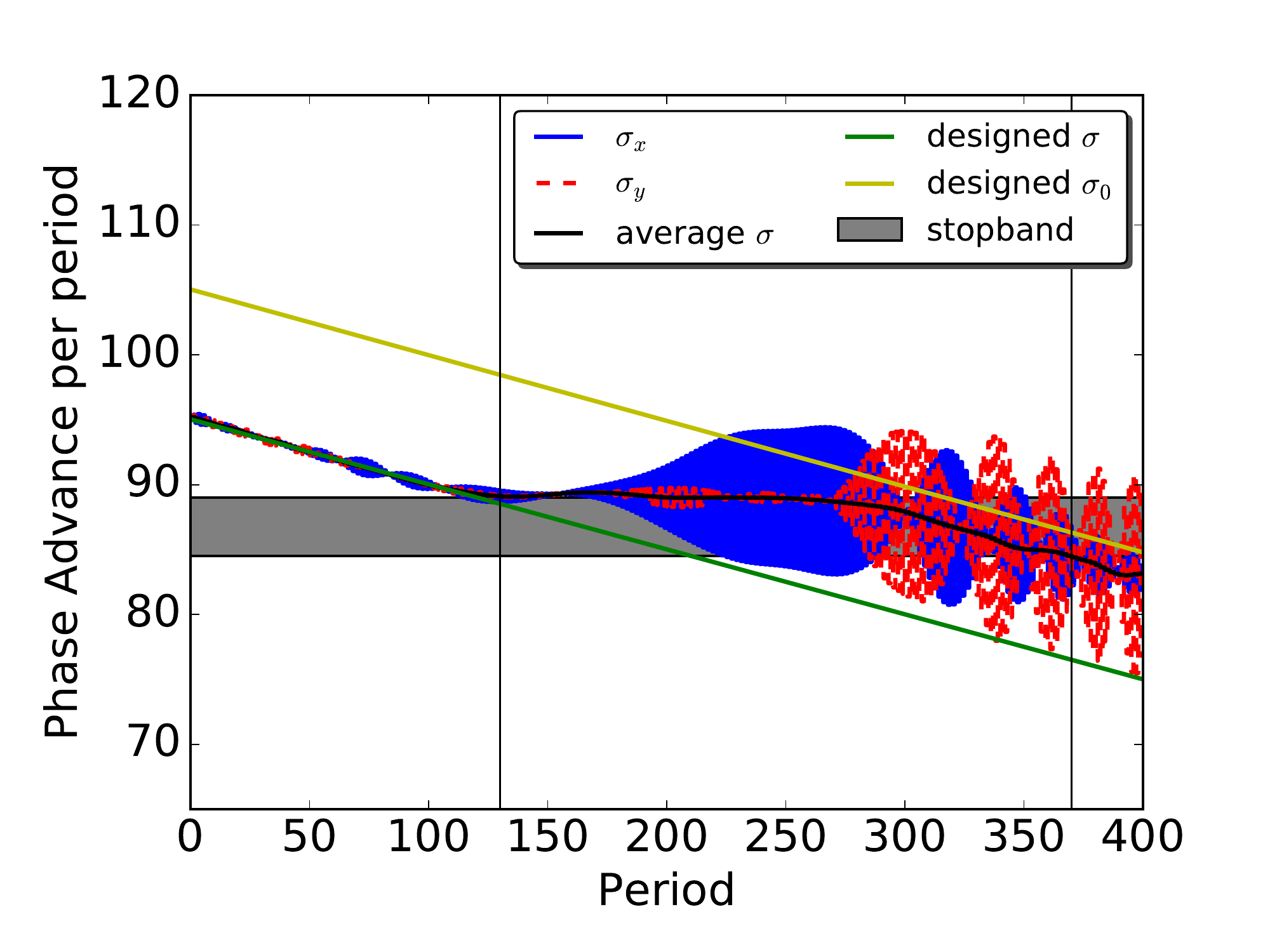}
    }
    \caption{
    Comparison of the phase advance evolution when the structure resonance
    region is crossed from below (left) and above (right). The red and blue
    curves are the periodic phase advance evolution in x and y direction
    obtained from simulation. The black curves are the center of the red and blue curves.
    The yellow and green curves are designed phase advance
    with and without beam current. The grey region (roughly
    $84^{\circ}<\sigma<89^{\circ}$) is the 2nd order structure resonance stop
    band.  The cross points between the
    designed (green line) and the averaged phase advance (black line) and the
    boundary of the stop band (grey region) indicate when beam gets into and
    out of the structure resonance stop band in the designed case and in the
    simulation. The vertical black lines indicate the locations where the
    beam gets into and out of resonance stop band in the simulation.}
    \label{fig:phase_advance}
\end{figure*}

\begin{figure*}[thbp]
    \centering
    \subfloat[\label{sfig:75_95emittance}]{
        \includegraphics[width=.45\linewidth]{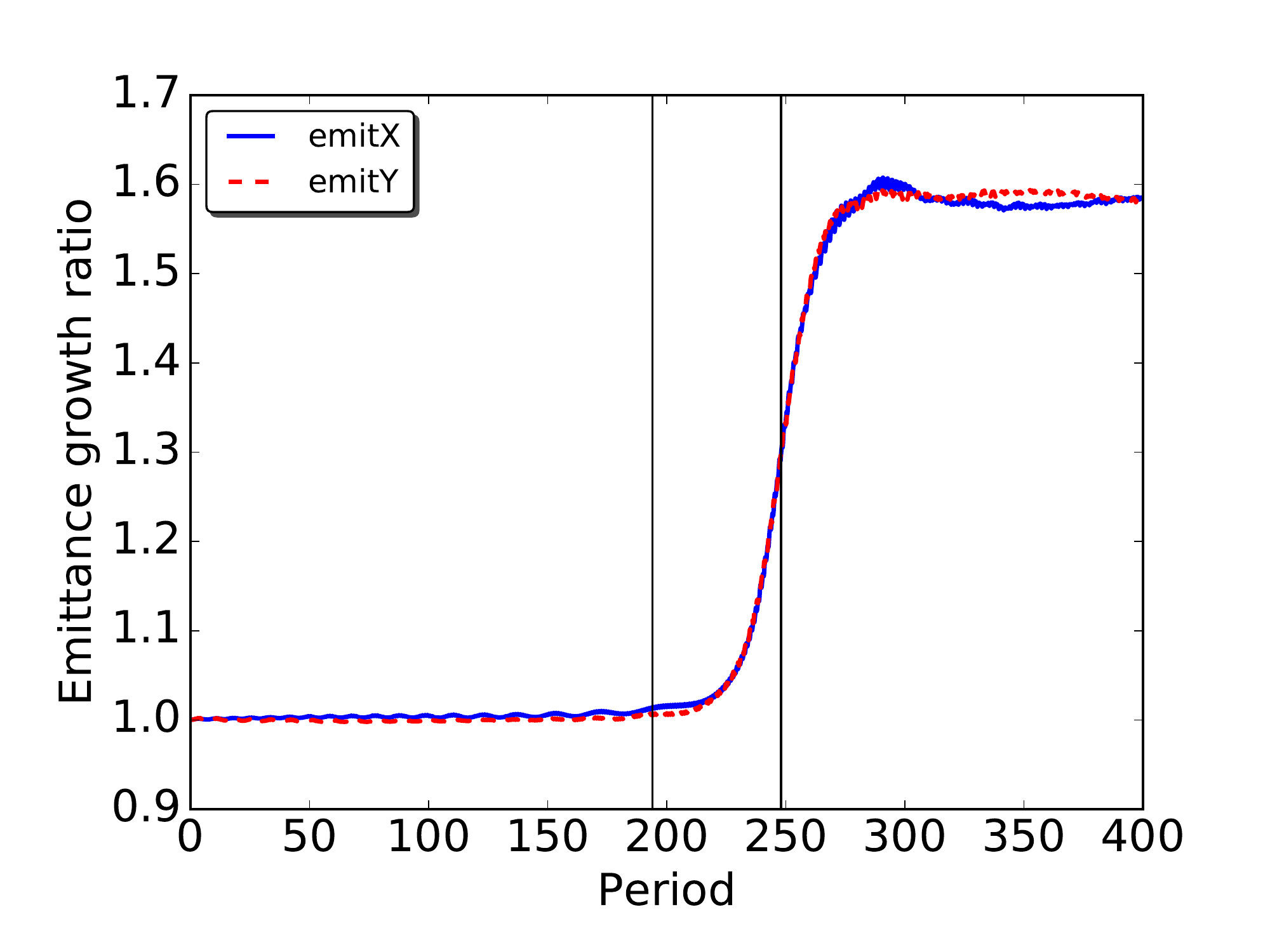}
    }
    \subfloat[\label{sfig:95_75emittance}]{
        \includegraphics[width=.45\linewidth]{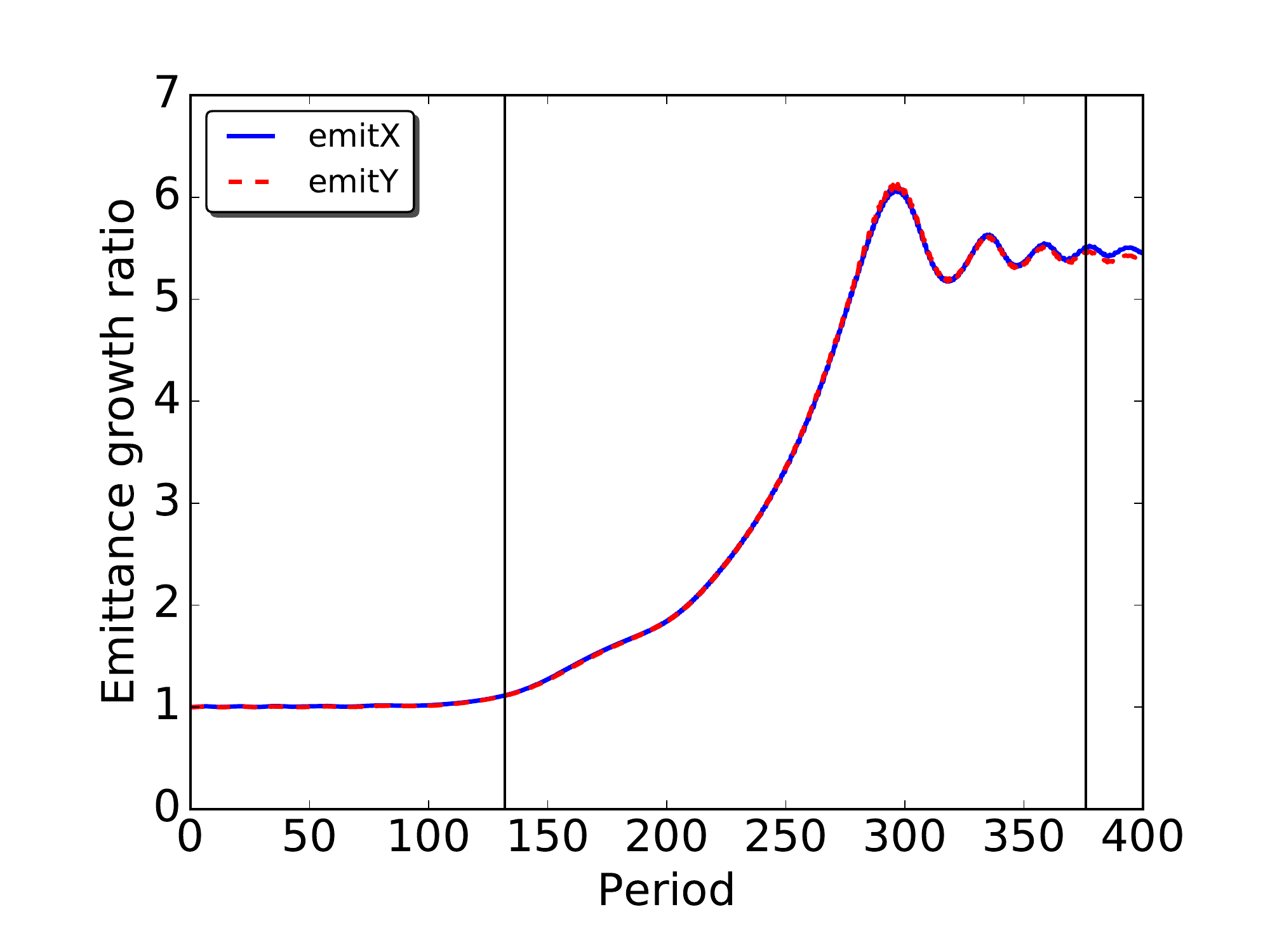}
    }
    \caption{Comparison of the rms emittance growth ratios
    ($\epsilon_f/\epsilon_i$) evolution when the structure resonance region is
    crossed from below (left) and above (right); The vertical black lines
    indicate the locations where the beam gets into and out of
    resonance stop band.}
    \label{fig:emittance}
\end{figure*}

\begin{figure*}
    \centering
    \subfloat[\label{sfig:75_95_ptc1}]{
        \includegraphics[width=.24\linewidth]{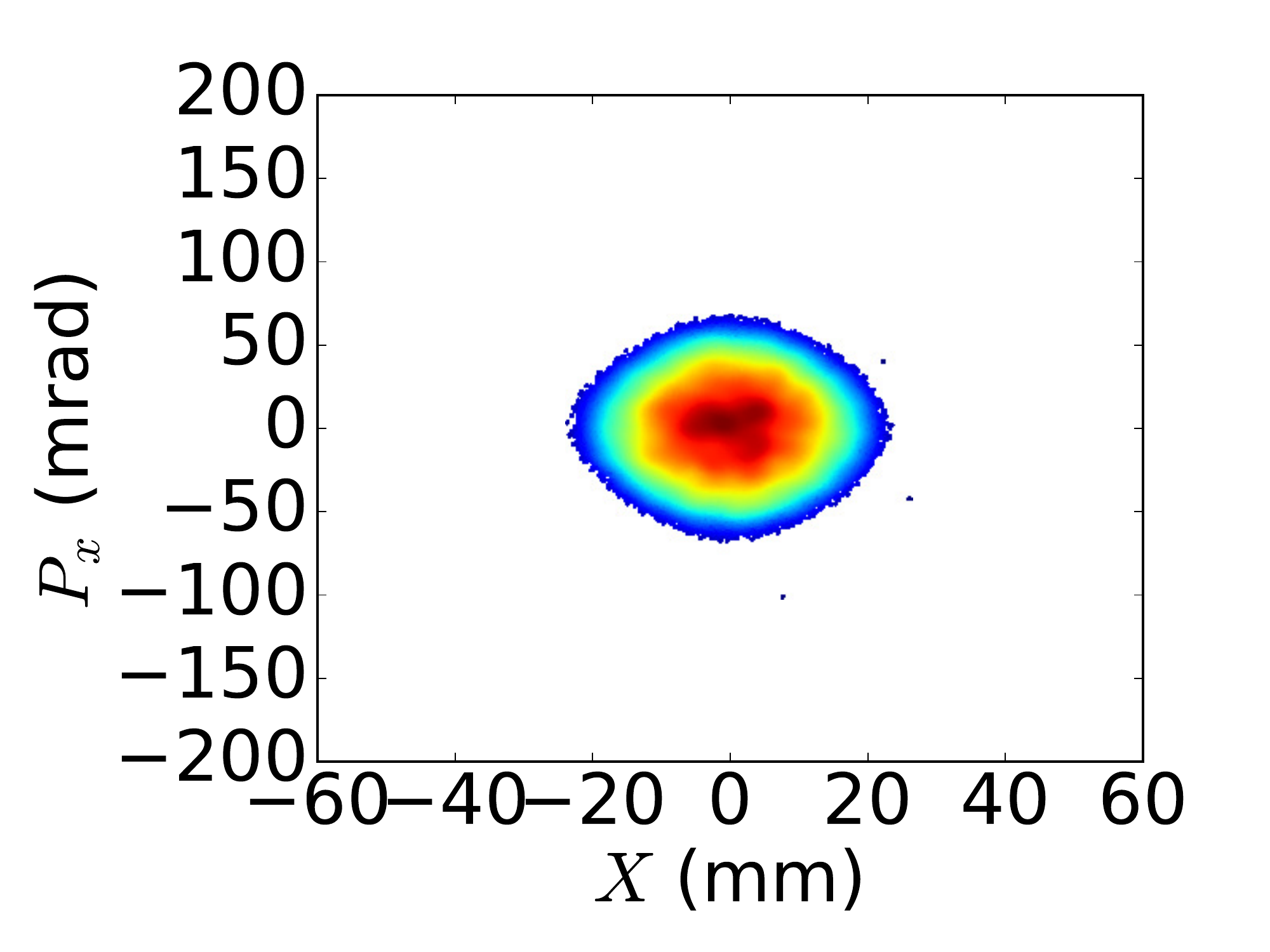}
    }
    \subfloat[\label{sfig:75_95_ptc2}]{
        \includegraphics[width=.24\linewidth]{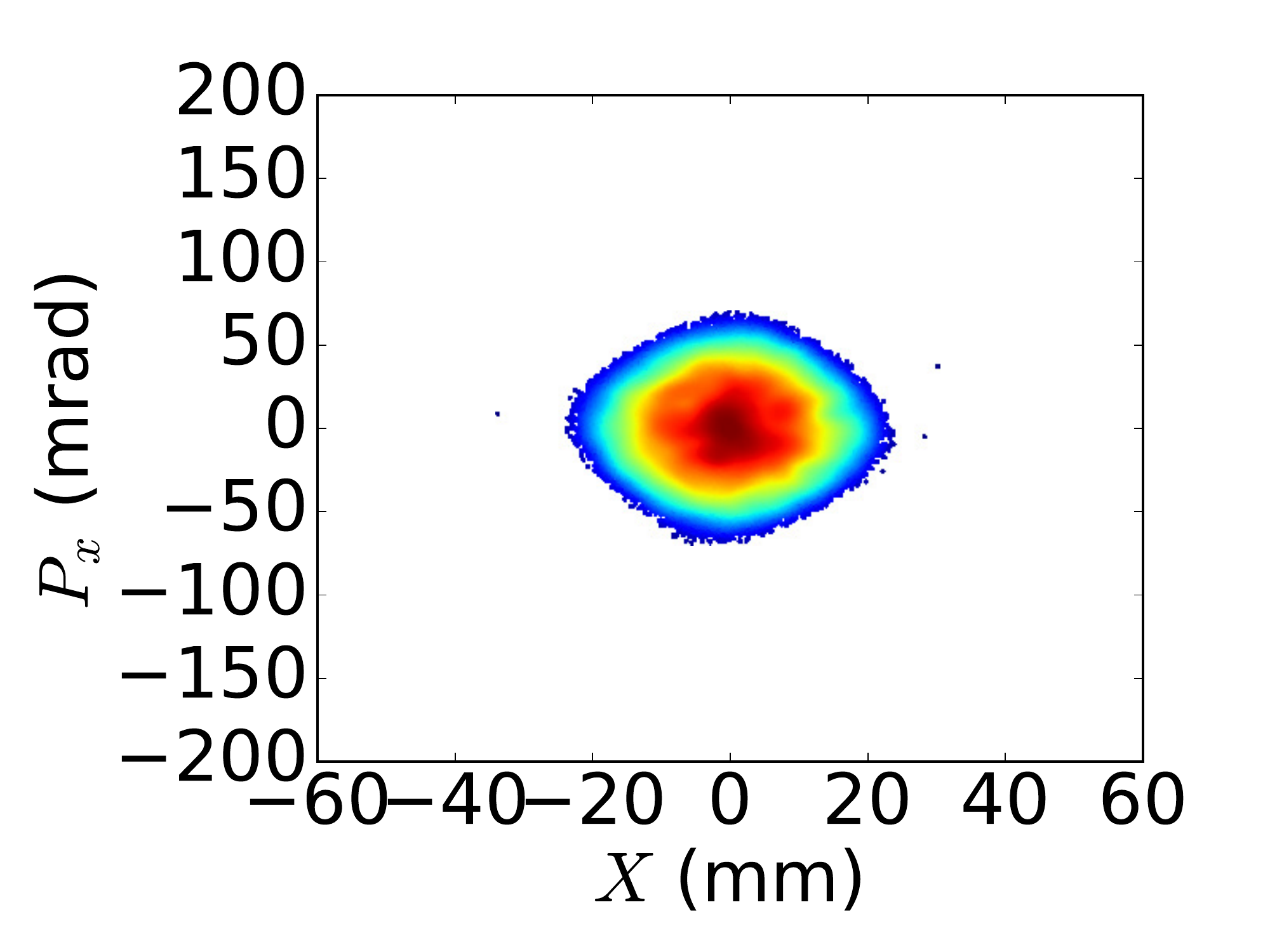}
    }
    \subfloat[\label{sfig:75_95_ptc3}]{
        \includegraphics[width=.24\linewidth]{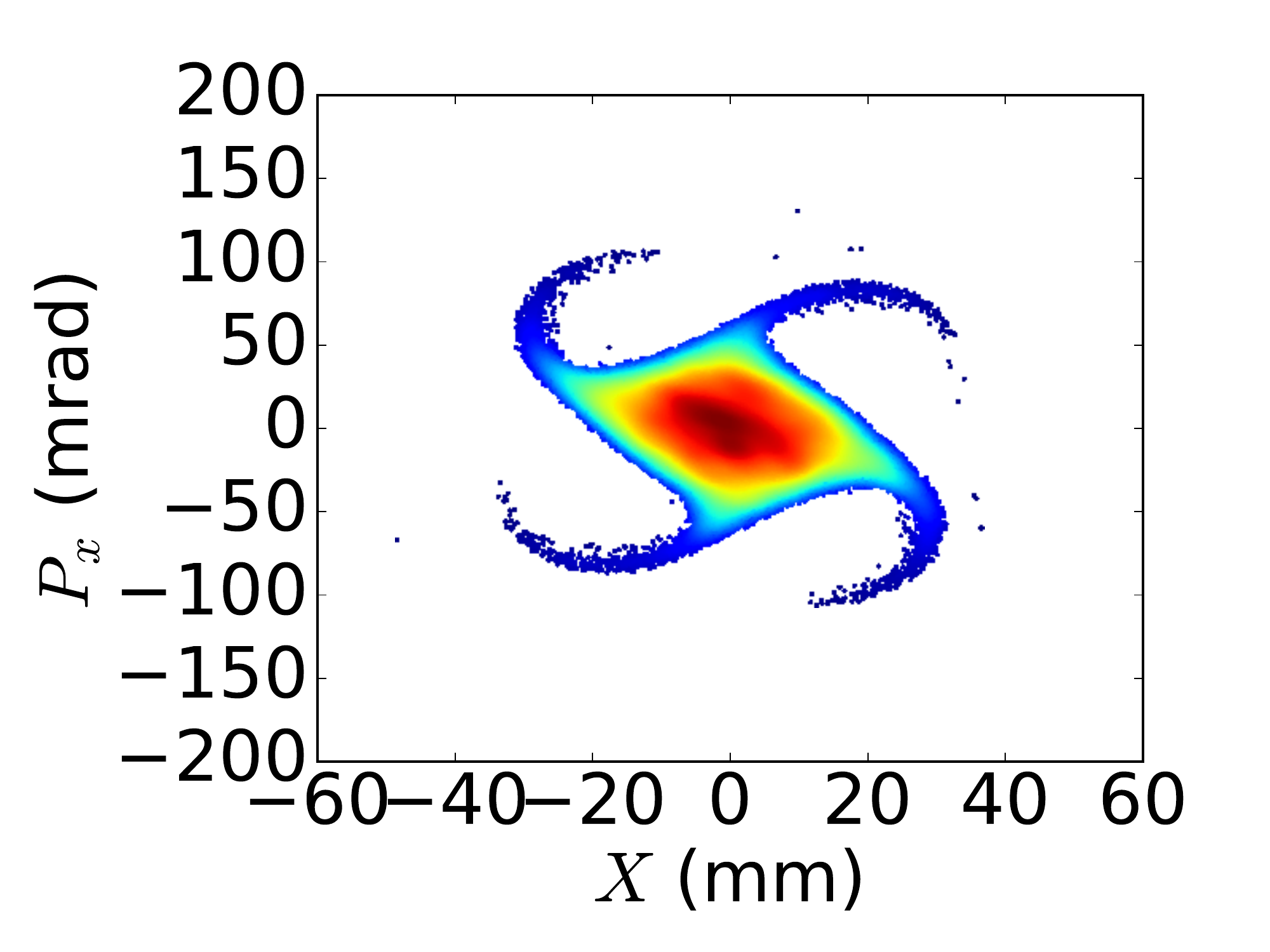}
    }
    \subfloat[\label{sfig:75_95_ptc4}]{
        \includegraphics[width=.24\linewidth]{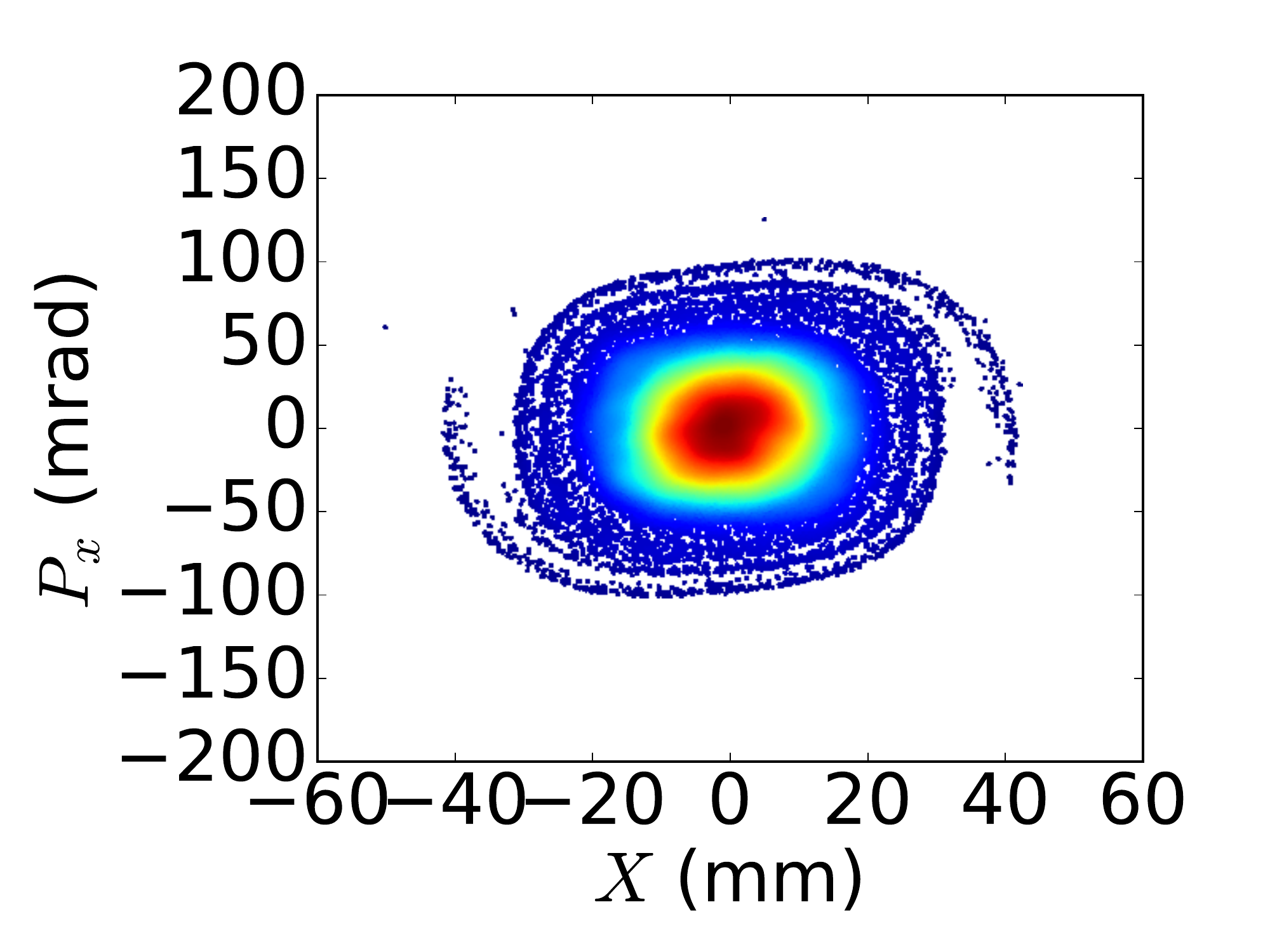}
    }

    \subfloat[\label{sfig:95_75_ptc1}]{
        \includegraphics[width=.24\linewidth]{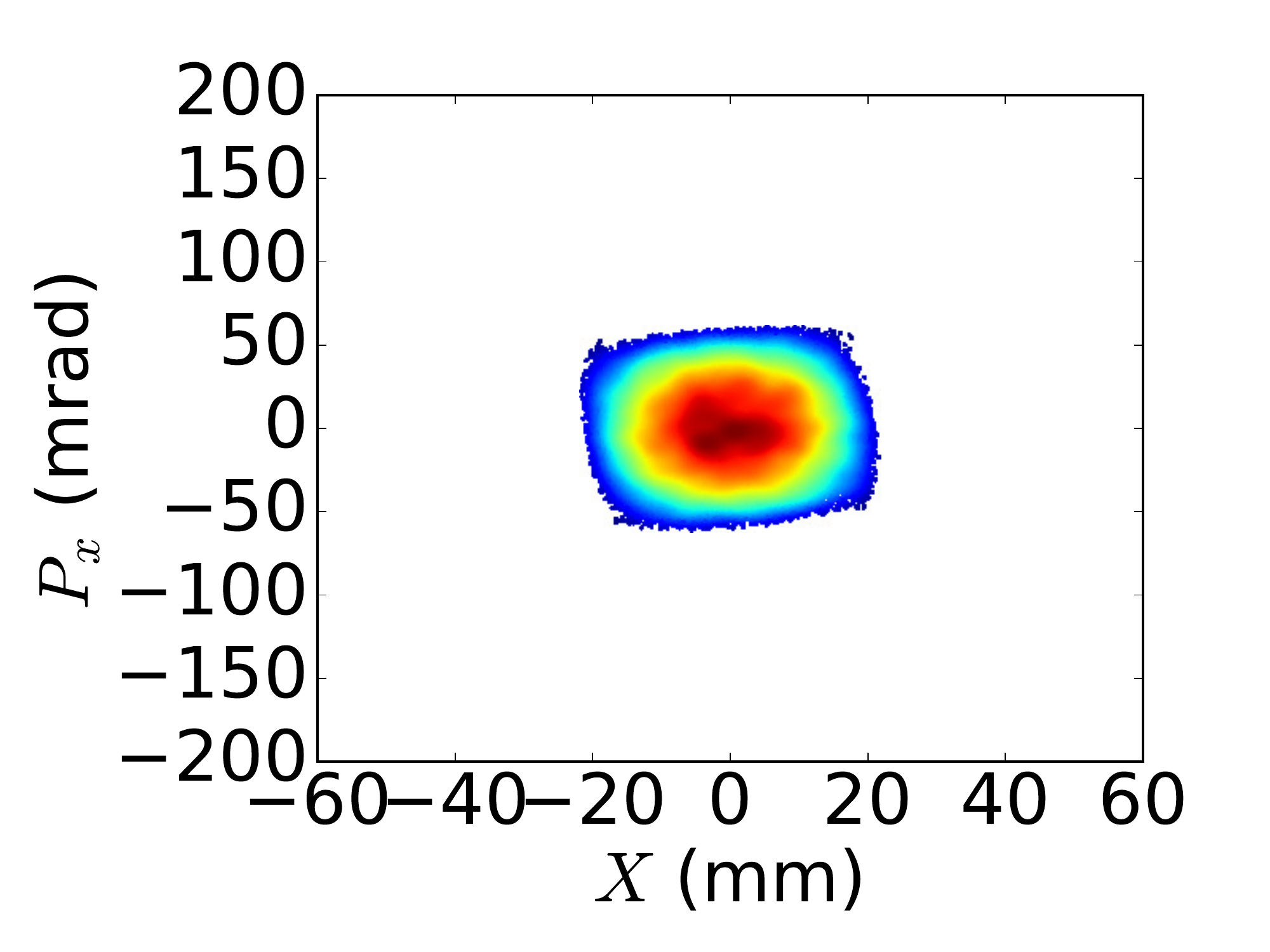}
    }
    \subfloat[\label{sfig:95_75_ptc2}]{
        \includegraphics[width=.24\linewidth]{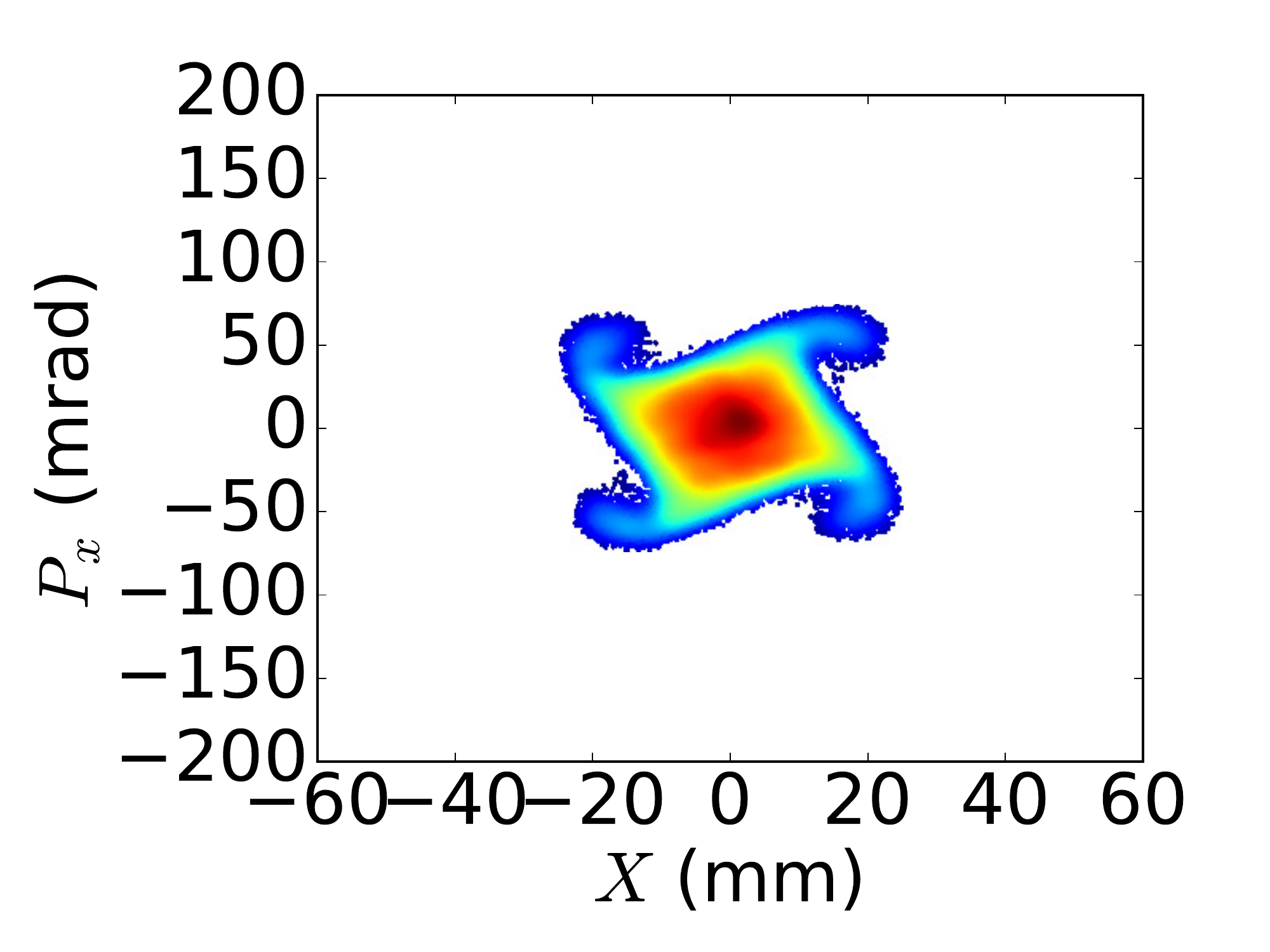}
    }
    \subfloat[\label{sfig:95_75_ptc3}]{
        \includegraphics[width=.24\linewidth]{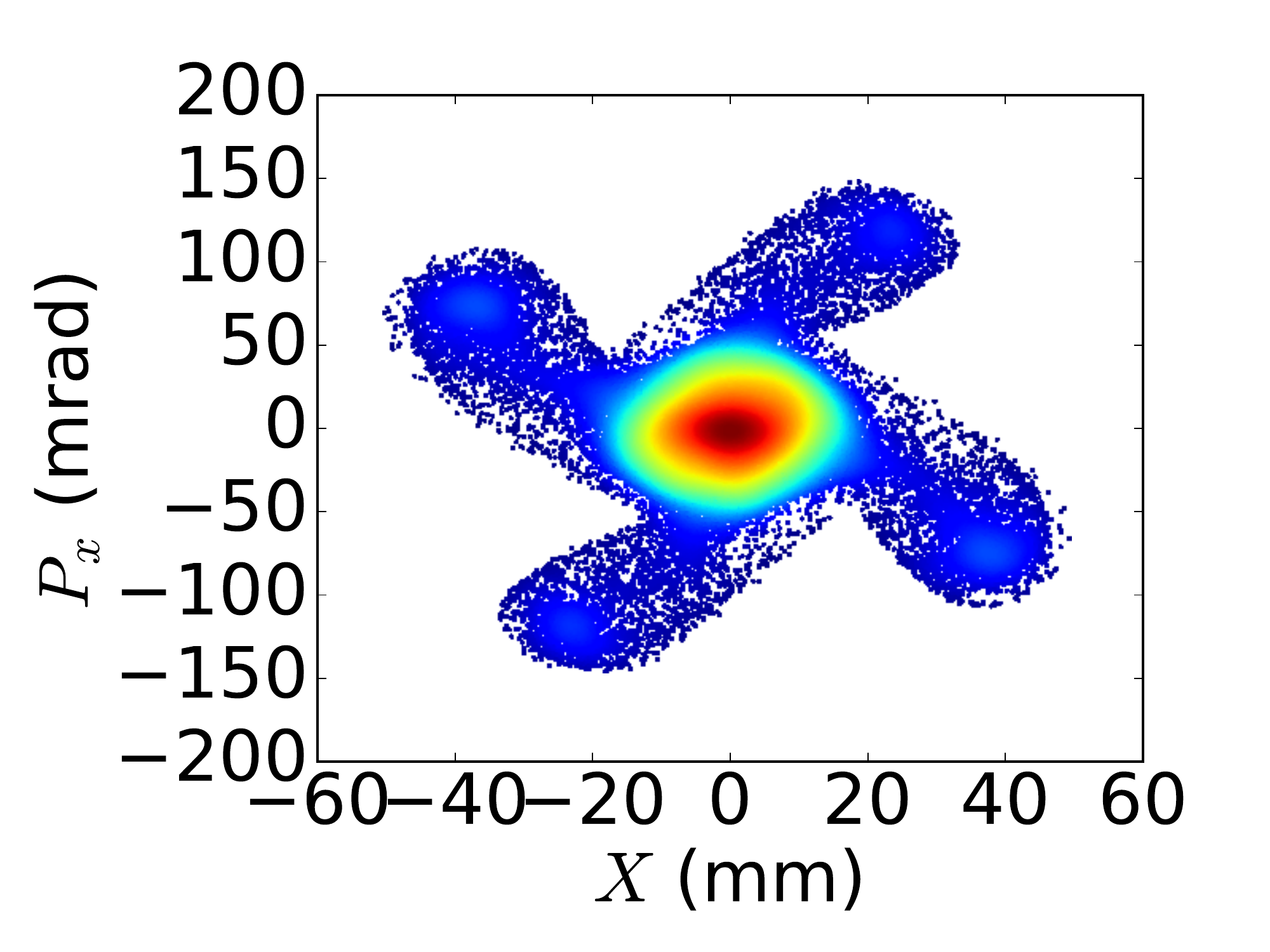}
    }
    \subfloat[\label{sfig:95_75_ptc4}]{
        \includegraphics[width=.24\linewidth]{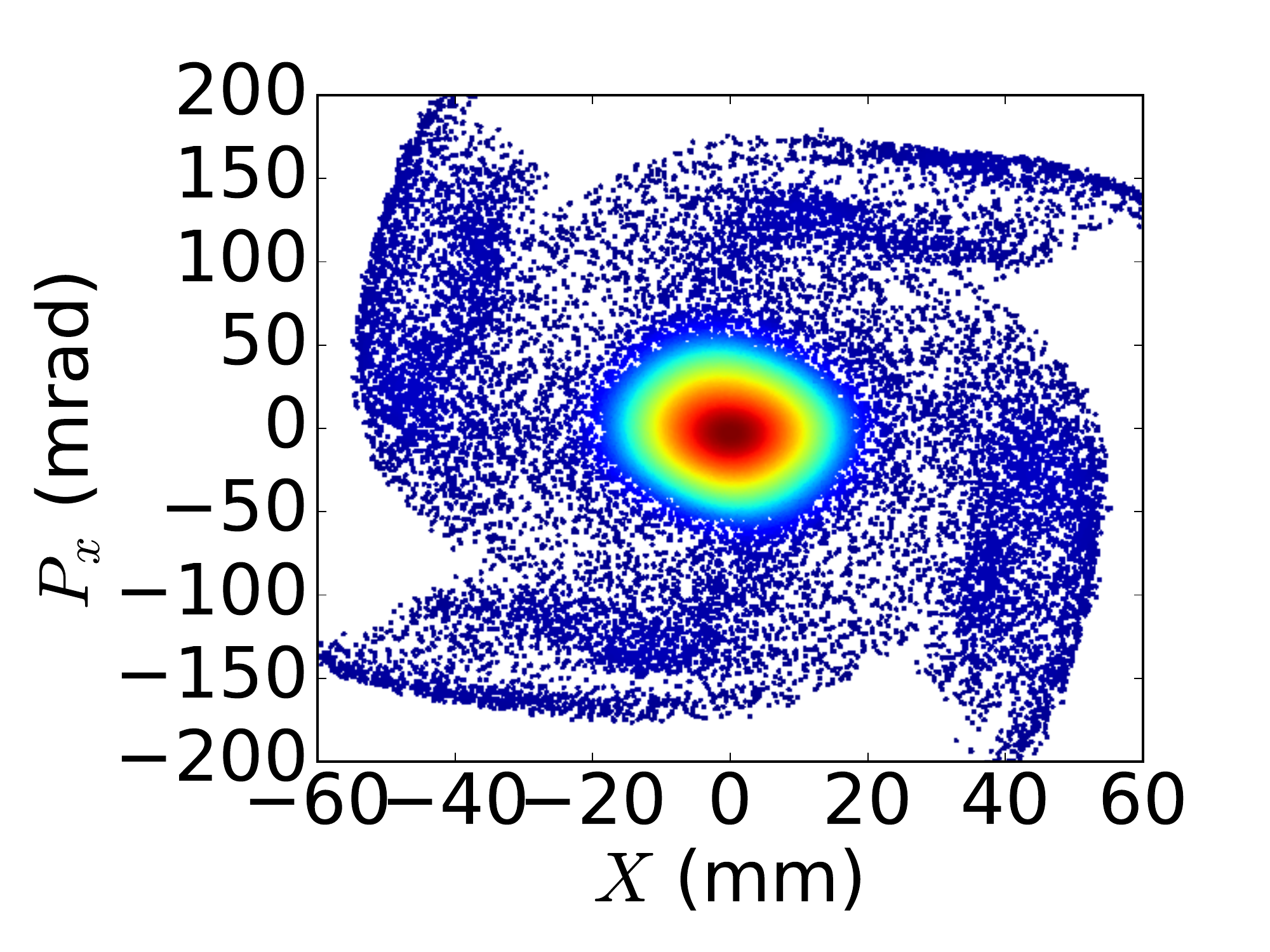}}
    \caption{The phase space $(x-p_x)$ profiles during structure resonance crossing from below (top) and above (bottom). The locations for each plot are chosen at period 50, 150, 250, 350.}
    \label{fig:95_75_ptc}
\end{figure*}

\begin{figure*}[thbp]
    \centering
    \subfloat[]
    {
        \includegraphics[width=.49\linewidth]{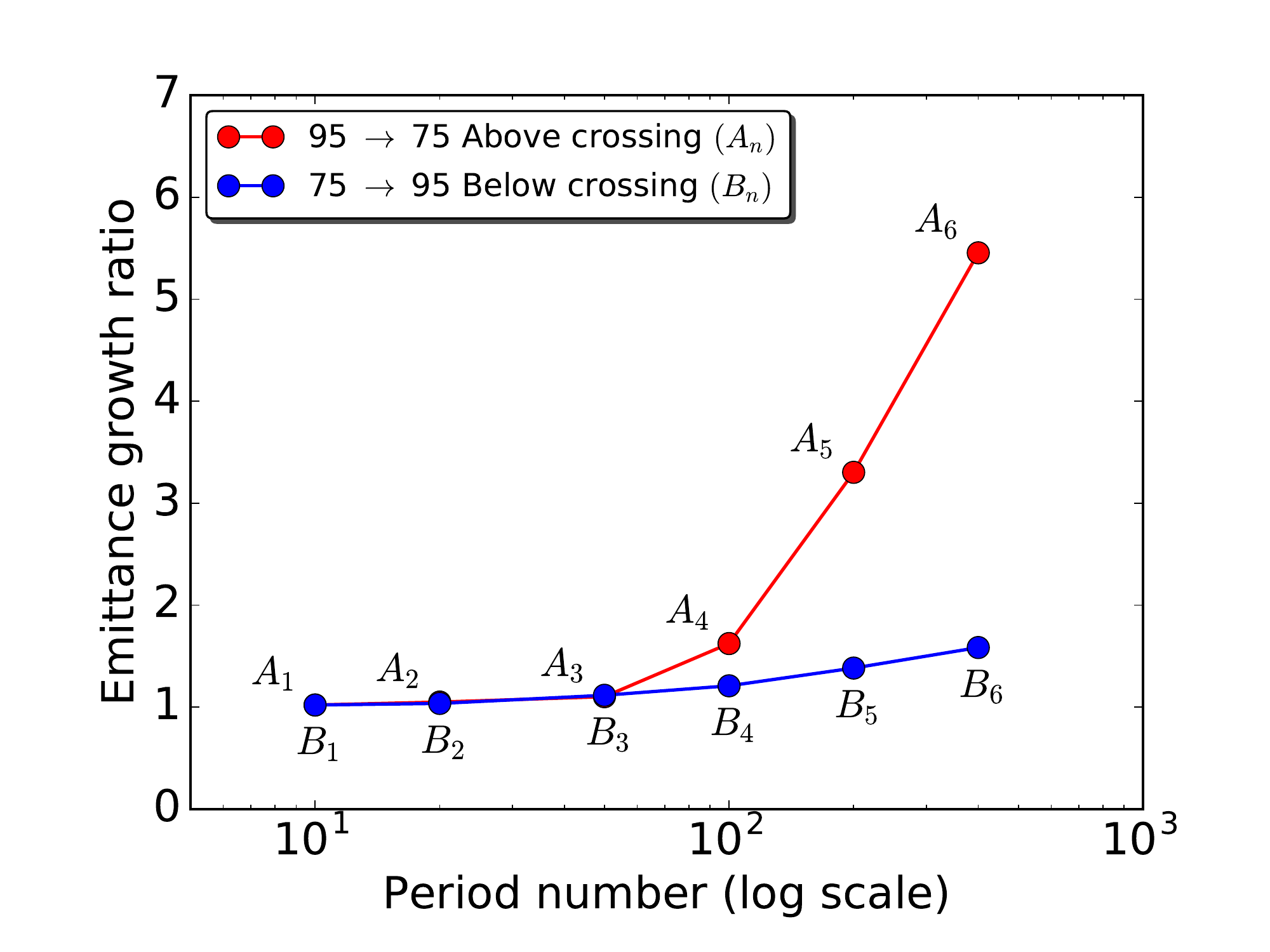}
        \label{sfig:CrossingSpeed_byStopband1}
    }
    \subfloat[]
    {
        \includegraphics[width=.49\linewidth]{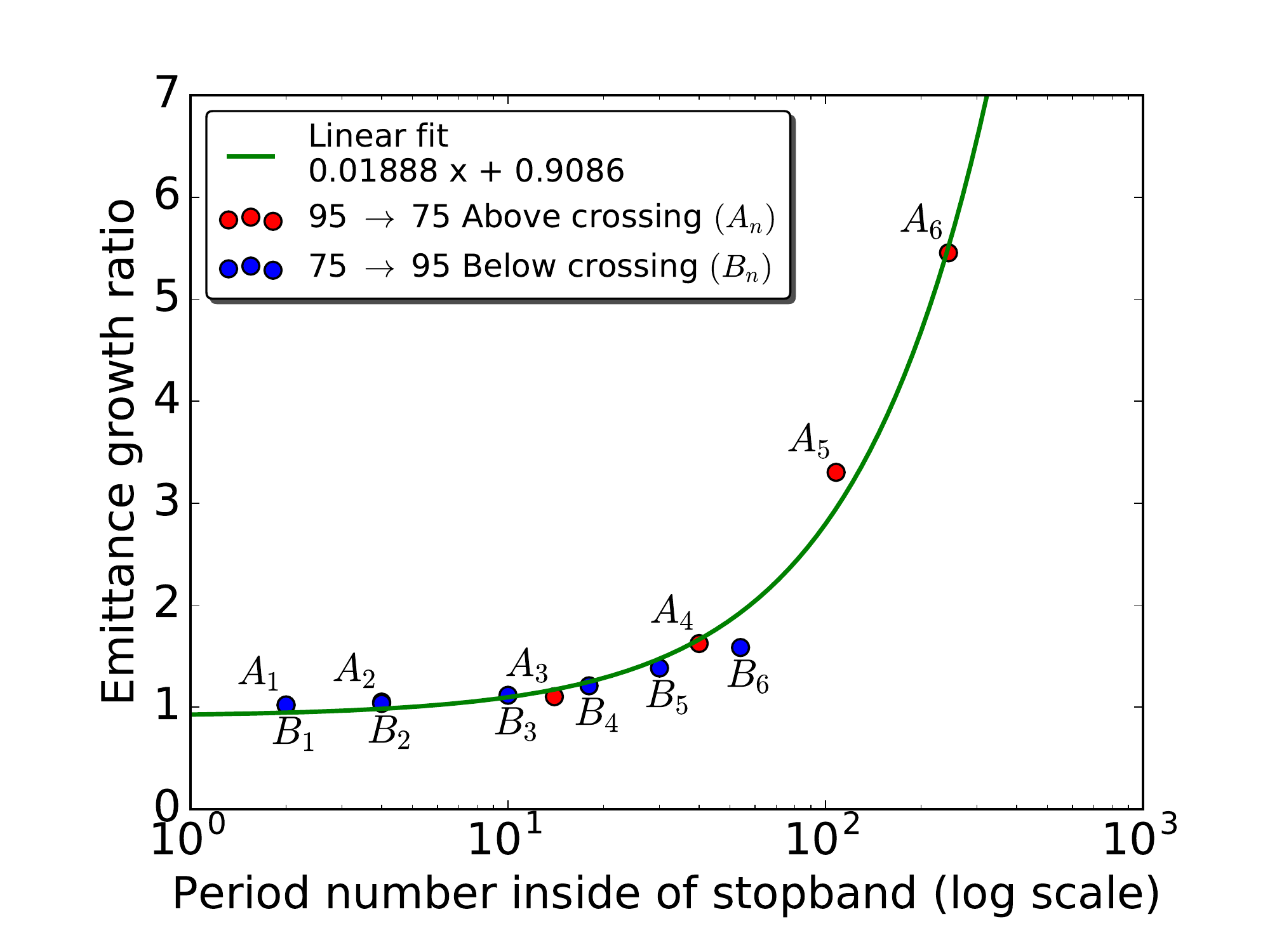}
        \label{sfig:CrossingSpeed_byStopband2}
    }
    \caption{
    a): The final emittance growth ratio after 400 FODO periods versus the
    number of FODO periods used in resonance crossing speed control within 400 FODO
    periods;
    b): The final emittance growth ratio after 400 FODO periods versus the
    effective number of periods where the beam is in the structure resonance
    stop band. The abscissas in above two figures are plotted on a logarithmic scale.}
    \label{fig:CrossingSpeed}
\end{figure*}

In the following, the Paritlce-In-Cell (PIC) code PTOPO \cite{23,24} is used for numerical simulations
to study the structure resonance crossing. The well matched WaterBag (WB) beam composed of 50000
macro-particles is used as initial beam distribution. Two approaches are applied for the rectangular meshes generation
in the Poisson solver. One is static case that the mesh size keeps the same as the aperture size; the other is the dynamics case that the mesh
size is as large as 6 times the rms beam size and varies  in each time step. 
In both cases, the Dirichlet boundary condition  is adopted.  
There is  no qualitative difference between the results of these two cases, and in the following study only the  
results from the  dynamical case are given.
The pipe size is large enough to ensure there is no beam loss during the whole calculation.
The lattice made up of 400 FODO periods with the focusing strength varying linearly,
which is long enough in cases discussed here to ensure no further emittance growth in simulations,
is used to imitate the structure resonance crossing process.

With a fixed beam current and the adiabatically varying quadrupole focusing
strength, two cases of the beam evolution are studied when the 2nd order
structure resonance stop is crossed. One is from the below (depressed phase
advance $\sigma$ varies from $75^{\circ}$ to $95^{\circ}$ during 400 periods
linearly), and the other is inversely from the above.
Fig.~\ref{fig:phase_advance} shows the evolution of phase advances along the
lattice when the structure resonance stop band -- grey region roughly with
$84^{\circ}<\sigma<89^{\circ}$ -- is crossed from below and above. The blue and
red curves represent the periodic phase advances in x and y direction respectively, which is obtained
with the help of the equation $\sigma=\int_{s_0}^{{s_0}+S} 1/\beta(s) ds  $. The black curves are the centers of the red and blue curves.
In principle, the simulated phase advances are limited in the region bounded by
the yellow and green lines which are the designed depressed phase advance $\sigma$ and
zero-current phase advance $\sigma_0$. Fig.~\ref{fig:emittance} shows the rms emittance
evolution correspondingly.

In Fig.~\ref{sfig:75_95phase}, when the structure resonance is crossed from
below, the simulated beam phase advance obeys the designed values spontaneously
until the beam gets into the stop band around period 200. Thereafter the speed
of beam structure resonance crossing gets ``faster'' than it is supposed to be
(indicated by the cross point between the green line and the boundary of the
stop band) and the beam gets out of the stop band round period 250, after which
it attains a local equilibrium state finally.
Fig.~\ref{sfig:75_95emittance} indicates that the rms emittance growth
takes place once the beam gets into the structure resonance stop band and no
further growth appears after passing through it, till period 265, which will be
explained later.
Inversely, Fig.~\ref{sfig:95_75phase} and  Fig.~\ref{sfig:95_75emittance} show
the phase advance and emittance evolution when the structure resonance
stop band is crossed from above.
In Fig.~\ref{sfig:95_75phase}, it can be verified that the beam stays for a
longer time than it is supposed to be in the stop band, which means the speed of resonance
crossing is ``slower''. The beam gets out of the resonance stop band at around
period 370. Compared with the case of structure resonance crossing from below,
the beam has a larger emittance growth finally.

The structure resonance crossing must be studied in a transient sense \cite{12}. 
The down-threshold and
up-threshold of the stop band vary with the beam rms emittance growing during the crossing. As a
result, in the case of the below crossing, the up-threshold moves to near
$90^{\circ}$, and beam roughly suffers 15 more periods from structure
resonance, until around period 265; Similarly, for the crossing from above, the
down-threshold also increases a little bit, and actually beam gets out of stop
band earlier than it is supposed to be, roughly at period 300. Another
characteristic of these two sets of studies is the ``faster" and ``slower"
resonance crossing speed. It seems that the resonance stop band has an
``attracting'' effect if the beam crosses from below and an ``repulsive'' effect
if it crosses from above. The reason is, in a general sense, under the comprehensive
influence of an external field and internal space charge field, if there is
any “instability” taking place, the beam always spontaneously tries to get rid
of this imbalance force which normally leads to rms emittance or beam halo
growth. This process can also be viewed as a kind of ``relief" from the energy
point of view. It is a natural result that the beam itself always tries to get
to a state where the space charge takes a weaker influence \cite{17}.
Here, it is reflected by the fact that the transient beam tune depression
$\eta=\sigma/\sigma_0$ turns to be larger than designed values once beam is
affected by structure resonance.

Fig.~\ref{fig:95_75_ptc} shows the phase space distribution evolution along
the FODO lattice during the structure resonance crossing from below (top) and
above (bottom). Clearly, the 4-fold phase space structures -- the evidence of the 4th order
structure resonance -- start to develop since the beam gets into the resonance
stop band both from the below crossing and above crossing. Question might be
asked why the 4th order structure resonance appears in the 2nd order structure
resonance stop bands. Actually, the reason is already explained above that the
lower order stop bands are components of the higher order structure resonance stop
bands \cite{li16collective, 12}. In a general sense, the appearance of different orders of
structure resonances requires appropriate driving force (Eq.~\ref{eq2.4}).
Starting form a WB initial distribution, the 2nd and 4th order perturbed
potential exists from the very beginning and will evolve self-consistently.
Thus, both the 2-fold and the 4-fold structure phase space are supposed to
appear in this 2nd order structure resonance stop band (Another example will be
shown in Sec.~\ref{section:incoherent}). Finally, the significant beam halo is formed due to these
mixed structure resonance effect.

To ensure  generality of the above understanding, various initial beam distribution,
like Parabolic distribution and truncated Gaussian distribution, are used for similar study. The patterns of emittance evolution, phase advance
evolution and phase space structure are similar with the case of initial WB beam discussed here since the rms equivalence \cite{13}.

\subsection{The structure resonance crossing speed and rms emittance growth}

\begin{figure*}
    \centering
    \subfloat[\label{sfig:TestPtc1}]{
        \includegraphics[width=.24\linewidth]{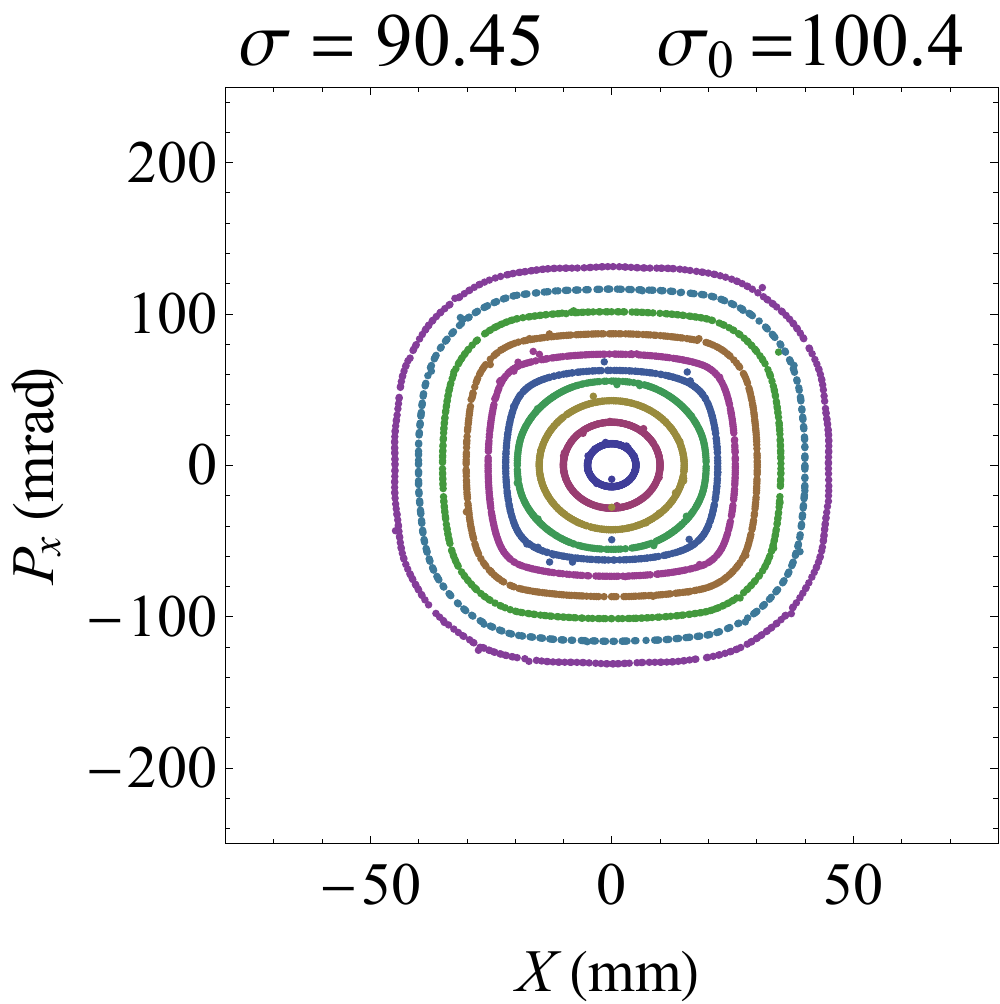}
    }
    \subfloat[\label{sfig:TestPtc2}]{
        \includegraphics[width=.24\linewidth]{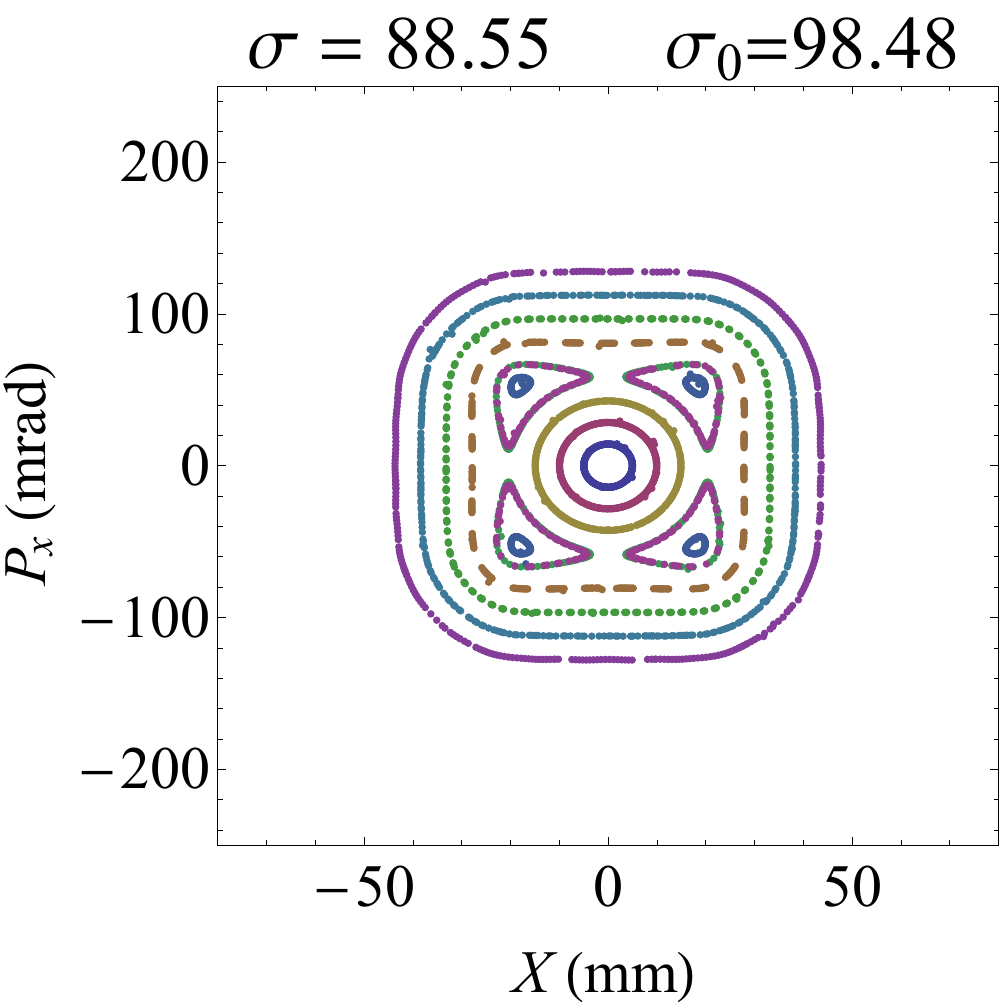}
    }
    \subfloat[\label{sfig:TestPtc3}]{
        \includegraphics[width=.24\linewidth]{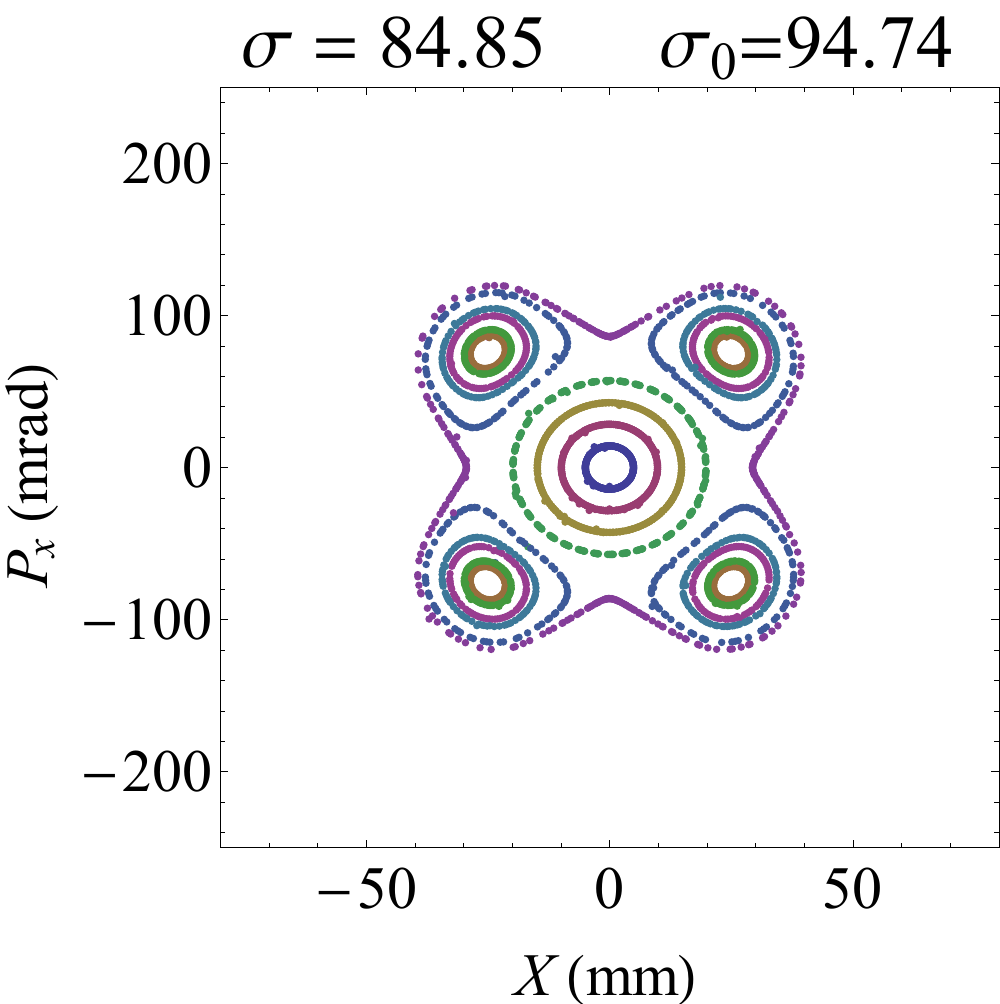}
    }
    \subfloat[\label{sfig:TestPtc4}]{
        \includegraphics[width=.24\linewidth]{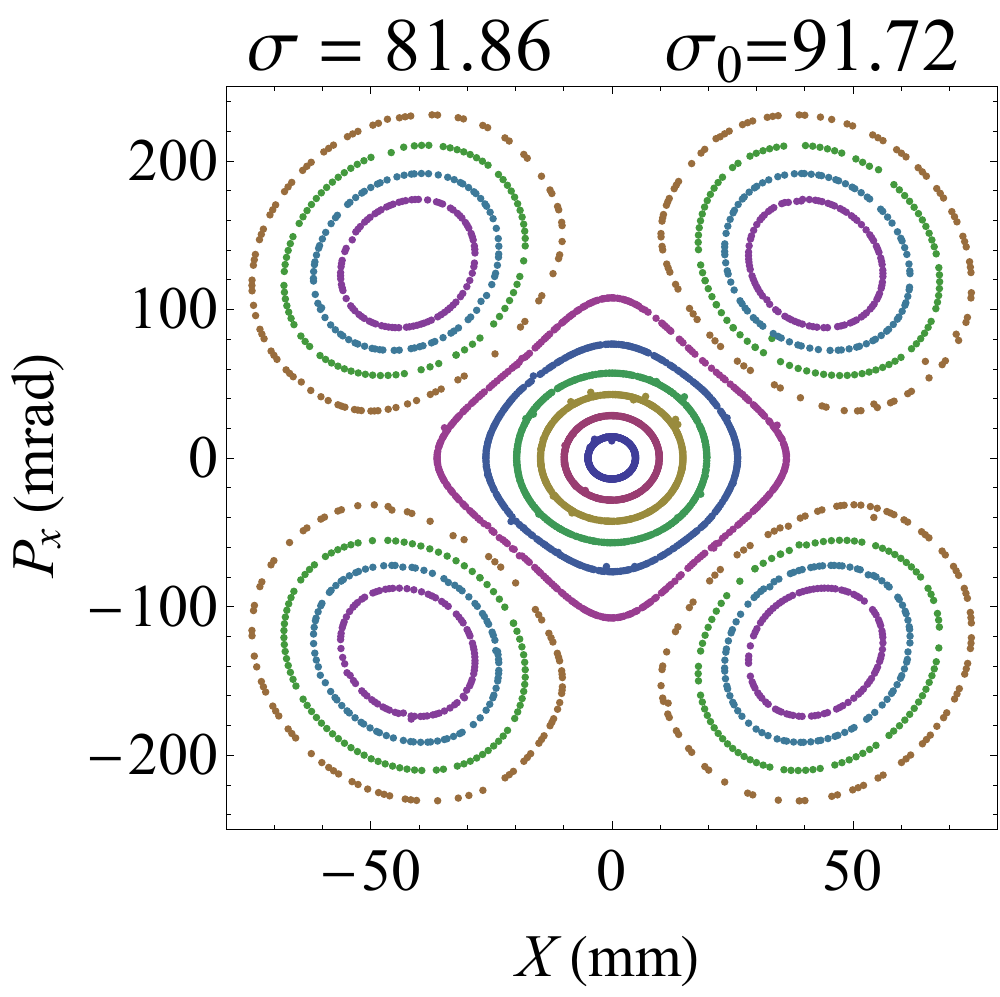}
    }
    \caption{Poinar\'{e} section from incoherent particle core resonance. Particle-core resonance islands move from inner core to outside in above crossing, Fig.~\ref{sfig:TestPtc1}$\rightarrow$Fig.~\ref{sfig:TestPtc4}, and conversely in below crossing Fig.~\ref{sfig:TestPtc4}$\rightarrow$Fig.~\ref{sfig:TestPtc1}.}
    \label{fig:TestPtc}
\end{figure*}

\begin{figure}[thbp]
    \centering
    \subfloat[\label{sfig:Ptc2tails1}]{
        \includegraphics[width=.49\linewidth]{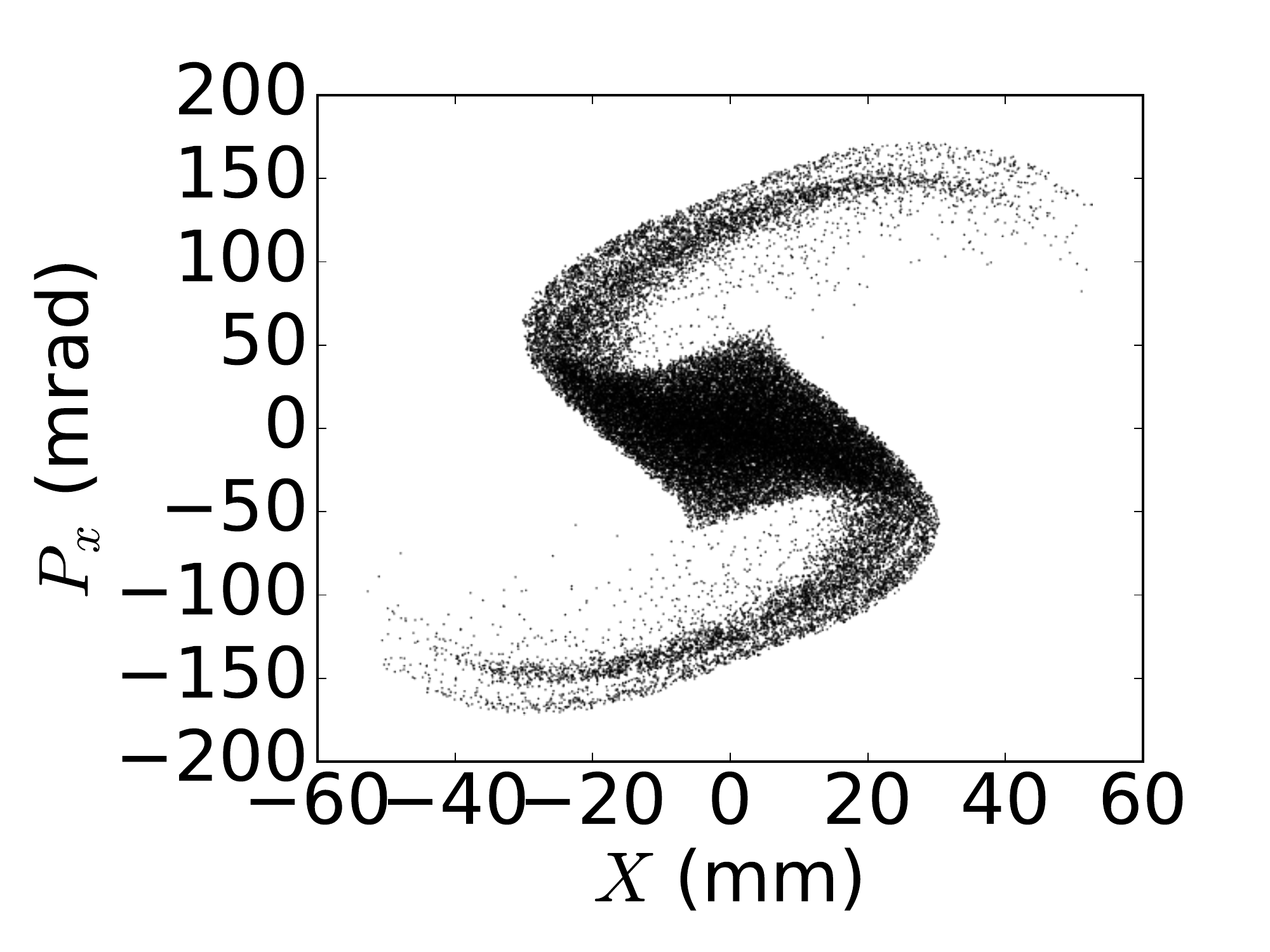}
    }
    \subfloat[\label{sfig:Ptc2tails2}]{
        \includegraphics[width=.49\linewidth]{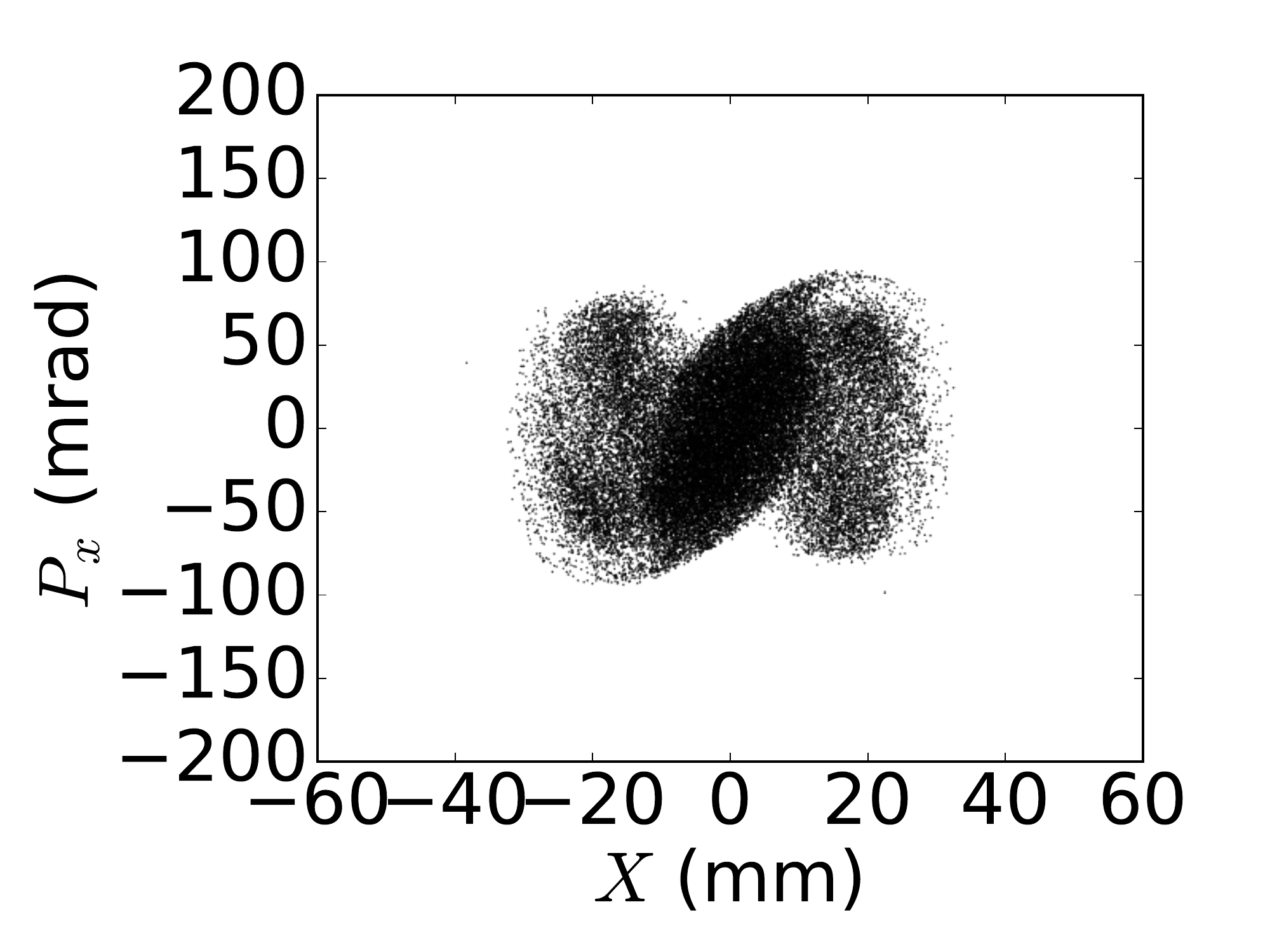}
    }
    \caption{Phase space profiles around a) period 205 in crossing from below, b) period 180 in crossing from above.}
    \label{fig:Ptc2tail}
\end{figure}

If resonance crossing cannot be avoided, it is now still under discussion
how it should be manipulated: cross it adiabatically or as fast as possible.
Here we mainly pay attention to the relationship between beam rms emittance and
the structure resonance crossing speed. With the same method to control the
beam phase advance as discussed in the above subsection, here, within total 400 FODO
periods, the first N (N=10, 20, 50, 100, 200, 400) periods are adjusted to make
the beam phase advances change linearly from $75^\circ$ to $95^\circ$ (below
crossing) and from $95^\circ$ to $75^\circ$ (above crossing). Again, the final
rms emittance growth after 400 periods is used as the evaluation of the beam quality.

Fig.~\ref{sfig:CrossingSpeed_byStopband1} shows the final emittance growth
after 400 FODO periods as function of the number of periods (N) used for
resonance crossing speed control.
The red and blue dots respectively represent the resonance crossings from above and below.
The different subscript $n$ represents different resonance crossing speed in which the focusing
strength are adjusted in the  first N (N=10, 20, 50, 100, 200, 400) periods.
With the same resonance crossing speed, the
emittance growth from the above crossing is much larger than that from the below
crossing, which indicates the same physics as we explained above in Fig.~\ref{fig:emittance}.
For different resonance crossing speed, both below crossing and above crossing
indicate that the structure resonance stop bands should be crossed as quickly
as possible to avoid significant emittance growth.
Fig.~\ref{sfig:CrossingSpeed_byStopband2} shows the emittance growth as a
function of the time that the beam spends within resonance stop band.
As expected, the emittance growth is
positively related to the time that the beam stays in the structure resonance
stop band (the horizontal coordinate in Fig.~\ref{fig:CrossingSpeed} is with logarithmic scale).

\section{Single particle dynamics -- incoherent effect}  \label{section:incoherent}

From the incoherent single particle dynamics point of view, it is intuitive to
attribute the 4-fold structure resonance to the 4th order particle-core
resonance ($90^\circ/360^\circ=1:4$) since the phase advance discussed in this
paper is mainly around $\sigma\sim90^\circ$.
Thus, the direction of the particle-core resonance islands moving towards or
outward to the core during structure resonance crossing is adopted to interpret
the 4-fold phase space structure \cite{25} and how severe of
beam halo when structure resonance stop band is crossed from below and above.

Using the particle-core resonance model~\cite{27,28}, Fig.~\ref{fig:TestPtc}
shows the Poincar\'{e} section plot of the single particle motion in phase
space. Clearly, when the structure resonance is crossed from above, four
particle-core resonance islands are generated in the origin point by a
bifurcation process \cite{29,30} and the formed islands move outward the core
during the structure resonance crossing process
(Fig.~\ref{sfig:TestPtc1} $\rightarrow$ Fig.~\ref{sfig:TestPtc4}).
In contrast, for the crossing from below, the particle-core resonance islands move
from outside into the inner core
(Fig.~\ref{sfig:TestPtc4}$\rightarrow$Fig.~\ref{sfig:TestPtc1}).
It is intuitive if one believes that the 4-fold phase space structure in phase
space is related to this particle-core resonance, since this particle-core
resonance can gradually bring particles in the core outside and form the stable
4-fold phase space structure; In another case, if particle core resonance
islands move from outside towards the inner core, since no particle is located
outside initially, less particles will occupy the 4-fold islands and less halo
particles are generated. However, this is misleading, because the fold
structures are mainly related to the coherent collective effect rather than the
incoherent particle-core resonance if the space charge is the only nonlinearity
source in the beam system. For instance, this incoherent particle-core
resonance cannot predict a two-fold phase space structure which actually
appears in the simulations of  both below crossing and above crossing, as shown in Fig.~\ref{fig:Ptc2tail}.
It is exactly the evidence of the 2nd order structure resonance -- coherent
effect. As pointed out by the former study \cite{17}, we argue that the
incoherent particle-core resonance might lead to the 4-fold phase space
structure, but only on a long-time scale. Here, we believe that the formed
4-fold phase space structure attained in the simulation corresponds to the mixed
4th/2nd collective structure resonance.

\section{Summary and acknowledgements}  \label{section:summary}
As a follow-up of the previous analytical study of the structure resonance \cite{li16collective,12},
this paper studies how the beam is spontaneously affected by the structure
resonance. Since the analytical studies are based on the KV beam distribution
assumption and the linearized perturbation theory, the nonlinear resonance damping effect
and the resonance saturation effect are not included.  The study in this paper
clearly shows the transient behavior of the beam when the structure resonance
stop band is crossed. The mixed 2nd/4th order coherent structure resonance
gives quite reasonable explanation to the results obtained from the PIC
simulations. It must be emphasized again that the interaction between the beam
and the resonance needs to be studied in a transient sense. The ``attracting''
and ``repulsive'' effect of the stop band is a spontaneous beam reaction since
the rms characteristics are modified by the structure resonance. It is also
found that the beam emittance growth is positively related to the
time that the beam is affected by the structure resonance. The final beam
equilibrium status is a comprehensive result as a comprise of the structure
resonance and the nonlinear resonance damping. The incoherent particle-core resonance
can cause phase space distortion, emittance growth, and beam halo formation,
but only on a long-time scale.

Another importance in this study is that the simulation clearly proves the conclusion from
previous theoretical prediction, that the lower order stop bands are
naturally included in higher order stop bands \cite{li16collective, 12}.  The study of higher order of
structure resonance can be  extended and understood in the same frame
discussed here.
One example is the 3rd/6th mixed structure resonance around $60^{\circ}$
phase advance $3 \times 60^{\circ} = 180^{\circ}$,
$6 \times 60^{\circ} = 360^{\circ}$[16] and $120^{\circ}$ phase advance
$3 \times 120^{\circ} = 360^{\circ}$, $6 \times 120^{\circ} = 720^{\circ}$
~\cite{li16collective,36}. The understanding of the coherent and incoherent resonance in
space charge physics leads us to a better understanding of recent experimental
results~\cite{31,32,33}. However, despite the progress in interpretation of the
nonlinear phenomena, there is still a large gap between analytical prediction, numerical simulation,
and the experiment result. In the near future, further study will be extended to the real 6D particle dynamics.

\begin{acknowledgments}
Great thanks for helpful discussion with R. A. Jameson in Institut f\"{u}r Angewandte Physik, Goethe Uni Frankfurt and Mei Bai in GSI. This work is supported by the Ministry of Science and Technology of China under Grant No. 2014CB845501.
\end{acknowledgments}

\appendix*
\section{The explicit form of the 2nd order Jacobi Matrix for collective structure resonance analysis.}
Note $C_{i,j}(s)=i/\beta_x(s)+j/\beta_y(s)$ for simplicity, where $\beta(s)$ is the betatron function, for the 2nd order even structure resonance:
\begin{eqnarray}\label{eq6.1}
J_{21} &=& - C_{2,0}^2 - C_{2,0} \frac{2 K}{\epsilon_x} \frac{a (2a+b) }{2  (a+b)^2} \nonumber  \\
J_{22} &=& \frac{C_{2,0}'}{C_{2,0}} \nonumber  \\
J_{23} &=& - C_{2,0} \frac{2 K}{\epsilon_x} \frac{a b }{2 (a+b)^2} \nonumber  \\
J_{41} &=& - C_{0,2} \frac{2 K}{\epsilon_x} \frac{a b }{2 \Gamma (a+b)^2} \nonumber  \\
J_{43} &=& - C_{0,2}^2 - C_{0,2} \frac{2 K}{\epsilon_x} \frac{ b(a+b) }{2 \Gamma (a+b)^2} \nonumber  \\
J_{44} &=& \frac{C_{0,2}'}{C_{0,2}};
\end{eqnarray}

For the 2nd order odd structure resonance,
\begin{eqnarray}\label{eq6.2}
J_{21} &=& - C_{1,1}^2 - C_{1,1} \frac{2 K}{\epsilon_x} \frac{a b (1+\Gamma)}{2 \Gamma (a+b)^2} \nonumber  \\
J_{22} &=& \frac{C_{1,1}'}{C_{1,1}} \nonumber  \\
J_{23} &=& - C_{1,1} \frac{2 K}{\epsilon_x} \frac{a b (-1+\Gamma)}{2 \Gamma (a+b)^2} \nonumber  \\
J_{41} &=& - C_{1,-1} \frac{2 K}{\epsilon_x} \frac{a b (1+\Gamma)}{2 \Gamma (a+b)^2} \nonumber  \\
J_{43} &=& - C_{1,-1}^2 - C_{1,-1} \frac{2 K}{\epsilon_x} \frac{a b (-1+\Gamma)}{2 \Gamma (a+b)^2} \nonumber  \\
J_{44} &=& -\frac{C_{1,-1}'}{C_{1,-1}}.
\end{eqnarray}
Here $K$ is the generalized  perveance,  $a(s)$ and $b(s)$ are the beam size during one period, $\Gamma=\epsilon_x/\epsilon_y$ is the emittance ratio between different degrees of freedom. In this paper, $\Gamma=1$ is selected for analysis.

\bibliography{reference}


\end{document}